# Thunderstorms, lightning, sprites and magnetospheric whistler-mode radio waves

Devendraa Siingh[1], A.K. Singh[2*], R.P. Patel[2], Rajesh Singh[3], R.P. Singh[2,4], B. Veenadhari[3] and M. Mukherjee[1]

[1]Indian Institute of Tropical Meteorology, Pune-411 008, India.
[2]Atmospheric Research Laboratory, Department of Physics, Banaras Hindu University, Varanasi- 221 005, India.
[3] Indian Institute of Geomagnetism, New Panvel, Navi Mumbai-410218, India.
[4] Vice-Chancellor, Veer Kunwar Singh University, Ara-802301 (Bihar), India.

**Abstract:**
Thunderstorms and the lightning that they produce are inherently interesting phenomena that have intrigued scientists and mankind in general for many years. The study of thunderstorms has rapidly advanced during the past century and many efforts have been made towards understanding lightning, thunderstorms and their consequences. Recent observations of optical phenomena above an active lightning discharge along with the availability of modern technology both for data collection and data analysis have renewed interest in the field of thunderstorms and their consequences in the biosphere. In this paper, we review the electrification processes of a thunderstorm, lightning processes and their association with global electric circuit and climate. The upward lightning discharge can cause sprites, elves, jets, etc. which are together called transient luminous events. Their morphological features and effects in the mesosphere are reviewed. The wide spectrum of electromagnetic waves generated during lightning discharges couple the lower atmosphere with the ionosphere/magnetosphere. Hence various features of these waves from ULF to VHF are reviewed with reference to recent results and their consequences are also briefly discussed.

**\* Corresponding Author**
Atmospheric Research Laboratory, Department of Physics, Banaras Hindu University, Varanasi- 221 005, India.
E-mail: abhay_s@rediffmail.com; devendraasiingh@tropmet.res.in
Fax No +91-0542-2368390

# 1. Introduction

A thunderstorm is characterized by strong winds in the form of squall, heavy precipitation and low level wind shear. The formation, intensification and propagation of thunderstorms are mostly governed by the synoptic and thermodynamic conditions of the atmosphere; their microphysical and electrical characteristics are known to affect significantly the formation and the intensity of precipitation. Thunderstorms are the deepest convective clouds caused by buoyancy forces set up initially by the solar heating of the Earth's surface. Several field and laboratory experiments have been conducted to determine the electrical nature of storms and possible electrification processes are being studied in the laboratory and also through theoretical modeling and computer simulations (Rycroft et al., 2007; Yair, 2008; Saunders, 2008, and references therein). Various research programmes such as the Thunderstorm Research International Programme (TRIP), the Down Under Doppler and Electricity Experiment (DUNDEE) (Rutledge et al., 1992), the Severe Thunderstorm Electrification and Precipitation Study (STEPS) (Lang et al., 2004) have been launched involving both ground and airborne measurements to study the electrical properties of thunderstorms and related phenomena.

The most fascinating aspect is lightning associated with thunderstorms whose strength and location can be assessed by a number of techniques such as those involving electrostatic, electromagnetic, acoustic, radar, and radio-frequency measurements. Electrostatic field/field-change measurements at multiple stations have the particular advantage of giving information about the electrical charge centers and their structure and movements in a thunderstorm (Krehbiel et al., 1979; Jacobson and Krider, 1976; Krehbiel et al., 2008). Usually electrified thunderstorms are modeled to have a dominant tripolar electrical structure, consisting of a negative charge in the middle and positive charges at the lower and the upper levels of the cloud, along with a negative screening charge layer at the upper boundary (Williams, 1989). However, the problem of determining the charge structure of storms from remote measurements of total electric field has some limitations, e.g. due to the vertical variation of conductivity and subsequent masking of upper charges. These problems are somewhat alleviated by measuring the time rate of change of electric fields, which forms the basis for measurements of the Maxwell currents (Krider and Musser, 1982). Such measurements can be used, in principle, to locate and quantify the different currents of the storm.

Thunderstorms exhibit cloud (including intra-cloud, cloud-to-cloud, and cloud-to-air), cloud-to-ground and cloud-to-ionosphere lightning discharges. Cloud-to-ground (CG) discharges are the most studied as a good part of them is observed from the Earth's surface. The discharges occur mostly between the main negative or positive charge center and the ground. Each flash consists of several strokes, with each stroke consisting of a leader and a return stroke; thus, negative or positive charges are brought to the ground. On the other hand, most of the intra-cloud (IC) discharges occur between the positive and negative charge centers of the main dipole. The upward discharges from cloud to the ionosphere may occur as a result of electrical breakdown between the upper storm charge and the screening charge attracted at the cloud top. They could also occur due to electrical breakdown between the main mid-level charge and a screening depleted upper-level charge that continues to propagate out of the top of the storm (Krehbiel et al., 2008). The first process has been used to explain blue jets, while the second one could explain gigantic jets (Krehbiel et al., 2008). Thus, recently observed optical emissions such as sprites, elves, jets, blue starters, etc. are associated with thunderstorms (Rodger, 1999; Barrington-Leigh et al., 2002; Su et al., 2003). Recent studies established a link between individual positive ground flashes that stimulated sprites and the excitation of global Schumann resonances within the Earth-ionosphere cavity (Boccippio et al., 1995).

Research in the past two decades has identified a surprising variety of "Transient Luminous Events" (TLEs). Amongst them the most common is the so called sprite, which is a manifestation of electrical breakdown of the mesosphere at 40 – 90 km altitude (Sentman et al., 1995). Sprites are associated with positive cloud to ground lightning discharges which lower positive charges from a cloud to the ground. Blue jets are discharges propagating upwards into the stratosphere from cloud tops in a similar way to classical lightning, consisting of leaders and a return stroke. They may or may not be associated with cloud to ground lightning activity (Wescott et al., 1995). Elves are concentric rings of optical emissions propagating horizontally outwards at the bottom edge of ionosphere at ~ 90 km altitude (Fukunishi et al., 1996), which are caused by the electromagnetic pulse radiated by the cloud to ground discharge current of cloud to ground lightning of either polarity (Cho and Rycroft, 1998).

Gigantic jets seem to be a discharge where a blue jet triggers a sprite, creating electrical breakdown of the atmosphere from the thunderstorm clouds directly up to the bottom of the ionosphere (Pasko et al., 2002; Su et al., 2003). Another related event is the Terrestrial Gamma-ray Flash (TGF) with energies up to 20 MeV, which is observed in association with lightning onboard a satellite (Fishman et al., 1994; Smith et al., 2005; Ostgaaard et al., 2008). TGFs may be bremsstrahlung radiation from upward propagating relativistic electron beams generated in a runaway discharge process powered by the transient electric field in the stratosphere and mesosphere following a lightning event. The runaway discharge process has also been suggested for the initiation of lightning and sprites (Roussel-Dupre and Gurevich, 1996), but so far no evidence of a direct connection between sprites and TGFs could be obtained.

Lightning discharges in thunderclouds radiate powerful radio noise bursts over a wide frequency range from a few Hz to several MHz. In the ELF/VLF frequency range waves can propagate over long distances in the Earth-ionosphere waveguide. Waves with frequencies less than 50 Hz can propagate globally with extremely low attenuation rates, allowing these radio waves to propagate a few times around the globe before dissipating. Interference between these waves results in the Earth-ionosphere cavity resonances known as Schumann resonances (SR) (Polk, 1982; Siingh et al., 2007). In this frequency range there are sources of interference due to electric railways, mechanical vibrations of the antennas, surrounding vegetation, drifting electrically charged clouds and power line transients, etc. Presently Schumann resonances are being used to monitor global lightning activity (Heckman et al., 1998; Rycroft et al, 2000), global variability of lightning activity (Satori, 1996; Nickolaenko et al., 1998) and sprite activity (Boccippio et al., 1995; Cummer et al., 1998a; Rycroft et al., 2000). The relations between lightning and ELF noise levels on the global basis have been used to study the space-time dynamics of world-wide lightning activity (Magunia, 1996). Schlegel and Fullekrug (1999) showed that solar proton events cause increases in the frequency, Q-factor and amplitude of the SR modes.

Waves in the very low frequency (VLF, 3-30 kHz) range penetrate the ionosphere and propagate along geomagnetic field lines without appreciable attenuation. Thus, the waves can propagate from one hemisphere to the other many times before being attenuated. The mode of propagation is called the whistler mode and the waves are termed whistlers. As whistler waves propagate from a lightning discharge to deep into the magnetosphere, they couple energy from the atmosphere to the magnetosphere. As the wave propagates through the ionized medium embedded in the geomagnetic field, it is dispersed and a particular form of whistler spectrogram (dynamic spectrum in the frequency-time domain) is obtained. Signals having spectra of different shapes are also observed and termed VLF emissions; these are broadly classified as hiss or unstructured emissions characterized by a continuous band limited signal producing a hissing sound (Helliwell, 1965), or chorus, structured emissions exhibiting coherent discrete spectra (Helliwell, 1969), and including periodic emissions, quasi-periodic emissions, triggered emissions (Helliwell, 1965).

Lightning discharges radiate intense electromagnetic pulses ~ 20 GW peak power for ~ 1 ms to ~ 1 s duration as measured by electric and magnetic sensors on the ground or in space (Neubert et al., 2008). The electromagnetic power heats the partly ionized layers of the upper atmosphere, the mesosphere and the D and E layers of the ionosphere. The quasi-static electric field of up to ~ 1 kV/m produced during a lightning discharge at mesospheric heights can accelerate electrons to relativistic energies; some of these might travel up into the magnetosphere. Whistler mode waves propagating along dipolar geomagnetic field lines interact with counter-streaming energetic electrons and scatter them from the Van Allen radiation belts into the atmosphere. These energetic electrons produce additional ionization in the D-region (Inan et al., 2007) and modify the electrical conductivity of the atmosphere (Hu et al., 1989). Inan et al. (1996) presented evidence of disturbances of electrical conductivity of the night time mesosphere and the lower ionosphere in association with lightning which lasted only 1 – 2 seconds. The resulting change in conductivity may cause changes in the amplitude/phase of VLF signals passing through the region. The most common perturbation observed follows within a few ms of the causative lightning discharge and has a short onset duration of less than ~ 50 ms; usually referred as early/fast (Inan et al., 1995). Recent observations show that early VLF amplitude perturbations for 90% cases were associated with + CG discharges which triggered sprites (Mika et al., 2005). On the other hand numerous + GG and – CG discharges which did not trigger sprite were seldom associated with amplitude perturbations. Thus sprites are nearly always accompanied by "early" VLF perturbations (Neubert et al., 2008). In a few cases "early" VLF perturbations were found to be associated with elves also (Neubert et al., 2008).

The precipitated (whistler wave induced) electrons (Rycroft, 1973) in the mesosphere and thermosphere through chemical effects can produce $NO_x$ (Rodger et al., 2007). Sprites may also produce $NO_x$ locally. The strong convection in a thunderstorm cell may carry tropospheric air into the lower stratosphere (Huntrieser et al., 2007). Recently efforts have been made to model the effect of sprite discharges on the chemistry of the middle atmosphere using different codes and observed properties of sprites (Enell et al., 2008; Sentman et al., 2008; Gordillio-Vazquez, 2008) The additional production of $NO_x$ affects the mesospheric ozone concentration and could perturb the pressure/temperature distribution of the stratosphere or change its dynamics.

The electromagnetic signals generated during lightning discharges carry information about the source region and the ambient medium through which they propagate. They can be used to estimate the ionization density in the magnetosphere, and the existence and location of plasmapause (Storey, 1953; Singh et al., 1998a). The electromagnetic energy radiated from sprite-associated lightning in the ELF range can be used as a diagnostic of sprites because optical methods become ineffective during daytime. The global nature of the SR phenomena provides special incentive to examine very distant events, at ~ 20 Mm distance (Williams et al., 2007a).

This introductory discussion clearly shows that cloud-to-ground lightning discharges and sprite discharges in the mesosphere have their origin in thunderstorms. In this review paper, we propose to briefly discuss the current status of our understanding of transient luminous events in the mesosphere, whistler mode signals in the ionosphere/magnetosphere, their parent-lightning discharges and thunderstorms. Thus, we uniquely try to summarize the phenomena taking place from the troposphere to the magnetosphere.

In section 2, a brief discussion about thunderstorm electrification is presented. Charge separation in a thunderstorm ultimately results in the lightning discharge which is also discussed in this section. Lightning discharges in thunderstorms are cloud-to-cloud, intra-cloud, downward (cloud-to-ground) and upward (cloud-to-ionosphere) discharges. In this review paper we attempt to discuss some phenomena associated with these discharges such as the global electric circuit, the role of lightning on processes affecting climate, and the effects of lightning on precipitation, aerosol, and cloud processes. In section 3, we briefly review recent work related to upward lightning such as sprites, elves and jets. In this section, we discuss the morphology, discharge mechanism and effects of these phenomena on stratospheric and mesospheric processes. In section 4 we briefly discuss electromagnetic very low frequency wave phenomena generated during electrical discharges in thunderstorms and different phenomena associated with wave propagation from the source region to the point of reception. The electromagnetic wave propagation through the ionospheric/ magnetospheric plasma and the development of plasma physics has led to many interesting results being obtained. However, we shall discuss only recent results. Finally, a summary and conclusions are presented section 5.

## 2. Thunderstorms

Thunderstorms generate and separate electrical charges whereas lightning discharges neutralize electrical charges. There are numerous processes operating synergistically within the environment of a mature convective cloud, and numerous processes with varying effectiveness and time-dependencies affect cloud electrification (Stolzenburg and Marshall, 2008; Yair, 2008). The charging of thunderstorms can be discussed as inductive charging or non-inductive charging. An inductive process requires pre-existing electric fields to induce charges on a particle so that when it rebounds from another charge is separated and the field enhanced. In the atmosphere the fair-weather electric field resulting from positive charges in the atmosphere and negative charges on the ground could be considered as the pre-existing field. Brooks and Saunders (1994) interpreted laboratory experiments to support this mechanism. However, there are experimental results from airborne instruments which require some other processes of charging. Non-inductive processes are independent of the presence of an external electric field. This process is based on collisions between graupel and cloud-ice particles and the selective transfer of charge of a certain polarity to the larger particle. In an ordinary thundercloud, the smaller ice crystals are charged positively and move upward, whereas larger graupel particles charged negatively descend relative to the smaller particles. This is the normal situation, depending on the prevailing conditions of temperature, liquid water content and mixing in the thunderstorm. A variant of this situation may lead to the reverse condition. Charging of solid particles can involve tribo-electric charging, charging by fracto-emission and photoelectric charging, as discussed in detail in different review papers (Yair, 2008; Saunders, 2008 and references therein). The presence of soluble ionic substances in the liquid and ice phases has a significant effect on the charging processes and may significantly alter the outcomes of particle interaction and the charging processes. In the non-inductive mechanism, Saunders (2008) has discussed drop break up, ion charging (atmospheric ions produced by cosmic rays, or radioactivity) and convective mechanisms, etc. He has also discussed particle charging involving the ice phase and ice crystal/graupel charging mechanism. However, he considers that the most viable processes are those by which ice particles, growing at different rates, collide and share charges such that particles growing fastest charge positively via the inductive mechanism.

The widely accepted model of thunderstorm electrification shown in Figure 1 is one in which conduction, displacement and precipitation current densities below the negative layer of charge in the thundercloud vary with altitude. (In the same figure are shown the various Transient Luminious Emissions (TLEs) of the stratosphere and mesosphere.) In the region between the bottom and the top of thunderstorm cloud charging, conduction, displacement and precipitation current densities vary in space and time. In the region above the thunderstorm only conduction and displacement currents are considered. All lightning currents are considered as discontinuous charge transfers. In the fair-weather regions far away from thunderstorms, only conduction currents flow.

The time variation of thunderstorm electric fields, both aloft and at the ground, can be interpreted in terms of total Maxwell current density ($J_M = J_E + J_C + J_L + \frac{\partial D}{\partial t}$) which varies slowly, where $J_E = \sigma E$ is the field dependent current, $J_C$ is the convection current produced by the mechanical transport of charges such as by the air motion or by precipitation, $J_L$ is the lightning current representing the impulsive, discontinuous transfer of charge in both space and time, and $\partial D/\partial t$ is the displacement current. In accordance with classical electrodynamics $J_E$, $J_C$ and $J_L$ are associated with charge transfer and can be considered as a part of the conduction current so that the total Maxwell current can be taken as the sum of the conduction current and the displacement current. The Maxwell current is almost unaffected during the evolution of thunderstorm when the electric field both at the ground and aloft undergoes large changes in amplitude, and sometimes even in polarity (Krider and Musser, 1982). They inferred that the cloud electrification processes may be substantially independent of the electric field, and that the thunderstorm is assumed to be a current generator. Considering ~ 1000 thunderstorms operating at any time and each generating ~ 1 Ampere from the top of the cloud charges the ionosphere with a charging current of $10^3$ A to a potential of ~ 250 – 300 kV (Rycroft et al., 2000; Williams, 2005; Tinsley et al., 2007). The ionosphere, being a good conductor, behaves as an equipotential surface having a potential of ~ 250 – 300 kV with respect to the Earth (Siingh et al., 2005a).

In the most simplified picture the thunderstorm is modeled as a dipole having charges at the bottom and top of the cloud and their image charges on the surface of the Earth, and the associated electrostatic fields determined (Farrell and Desch, 1992). A tripolar model for the charge structure of the thunderstorm has also been widely used (Williams, 1989; Krehbiel et al., 2008). In realistic cases charge distributions in thunderstorms sometimes may involve up to six charge layers in the vertical direction (Marshal and Rust, 1993; Shepherd et al., 1996). When computing the electric field, each of the charge centers can be viewed as generating its own polarization charge in and above the thundercloud and the resultant configuration of the electric field and charge density can be obtained using the principle of superposition.

**2.1 Spatial and temporal distribution of thunderstorm**

Lightning activity is concentrated in three distinct zones - East Asia, Central Africa and America - and is more prevalent in the northern hemisphere than the southern hemisphere mostly occurring over the land surface. The observation of lightning activity from space shows that two out of every three lightning flashes occur in tropical regions (Williams, 1992). In addition to the tropical lightning, extra-tropical lightning activity plays a significant role in the summer season in the northern hemisphere, resulting in the global lightning activity having a maximum from June to August. The similarity of the diurnal variations of the electric field over the oceans and of the worldwide thunderstorm activity supports the hypothesis that thunderstorms are the main electrical generators in the DC global electric circuit (GEC). The largest of the three maxima occurs at the time of maximum thunderstorm activity over the Americas, although this is weaker than that over Africa. Williams and Satori (2004) and Williams (2008) explain this as being due to greater importance of electrified shower clouds in driving the global electric circuit (also see Rycroft et al. (2007)) than thunderstorms in the Americas, where it is rainier than in Africa. This paradoxical effect could also be explained by the fact that South American thunderstorms are close to the magnetic dip equator, whereas most African thunderstorms occur over the Congo at higher (southern) dip latitude (Kartalev et al., 2004, 2006).

About 2000 thunderstorms are active at any time. These are mainly concentrated over the tropical land masses during the local afternoons and cover about 10% of the Earth's surface (Markson, 1978). Over the remaining 90% of the Earth's surface the return current (fair weather current) of ~1000A (~ 1 pA/m$^2$) flows from the ionosphere to the Earth's surface. There is also a good relation between the AC component of the global circuit (Schumann resonance) and global lightning activity (Clayton and Polk, 1977). All these studies show that the global circuit has a maximum current at ~1800 UT and a minimum at 0300 UT (Price, 1993).

Recently Sato et al. (2008) studied the temporal and regional variation of lightning occurrence and their relation to sprite activity and climate variability based on 1–100 Hz ELF magnetic field data obtained at the Syowa (Antarctica), Onagawa (Japan) and Esrange (Sweden) for the period from September 2003 to August 2004. They found that in the northern summer season (June to August) the lightning occurrence rates are higher in the northern hemisphere than in the southern hemisphere with large enhancements in North America, South-East Asia and the northern part of Africa. On the other hand, in the northern winter season (December to February) these rates are higher in the southern hemisphere, with large enhancements in South America, Australia and the southern part of Africa.

**2.2 Thunderstorms and Lightning**

In recent years some progress has been made towards understanding thunderstorm and lightning (Mazur and Ruhnke, 1993; 1998). To explain lightning phenomena, Kasemir (1950, 1960, 1983) introduced the concept of the bidirectional uncharged leader, emphasizing that the essential factor in maintaining the lightning discharge is the continuing breakdown at the tip of the positive or negative ends of the lightning leader that extends the channel into new regions with stored electrostatic energy.

General models of the cloud charge distribution are based on electric field measurements inside a thundercloud and on the ground (Stolzenburg, 1994 and references therein). A widely used model is based on the tripolar charge structure consisting of a negative charge in the middle of the cloud, a positive charge above it and a smaller positive charge below. Sometimes a screening negative charge on the upper cloud boundary is also considered (Krehbiel, 1986; Stolzenburg, 1994). A similar kinematic numerical model of thunderstorm electrification has been proposed by Ziegler and MacGorman (1994). Using this tripole model, Mazur and Ruhnke (1998) simulated the lightning discharge and showed that the upper part of the cloud-to-ground leader and the lower part of the intra-cloud leader terminate inside the cloud.

The charging current leads to charge build-up in the thunderstorm until a breakdown threshold is reached. At this point, bidirectional discharges are initiated, producing different types of lightning. When breakdown occurs between adjacent unbalanced charge regions, then discharges escape the storm. For example, breakdown triggered between mid-level negative charges and lower positive charges escape the storm downward to become a negative cloud-to-ground discharge. The charge imbalance (negative charge being much larger) imparts a strongly negative initial potential to the downward developing leader and, as such, the initiated discharge does not terminate at the lower positive charges.

The negative cloud-to-ground discharge transfers intermittently negative charge to the ground, thereby helping to charge the GEC and shifting the storm's net charge from negative to positive. The cloud potential quickly shifts to positive values and the electric field is enhanced in the upper part of the storm. Continued charging could initiate a discharge in the upper part of the storm within a few seconds, which may lead to an upward discharge having the same polarity as the upper storm charge, i.e. for a negative cloud-to-ground discharge, the upward discharge would have a positive polarity. The triggering is suppressed if the screening charge is mixed into the upper storm charge (Krehbiel et al., 2008). The infrequent observation of jets implies that mixing of the screening charge is normally strong in a storm. The upper level discharge, once triggered, would propagate upward above the cloud top and the distance covered would depend upon the positive potential in the upper part of the storm which is imparted to the developing leader channel. The upward discharge also helps to charge the GEC.

**2.2.1 Thunderstorms and the Global Electric Circuit**

The global electric circuit (GEC) links the electric field and current flowing in the lower atmosphere, ionosphere and magnetosphere forming a giant spherical condenser (Rycroft et al., 2000; Siingh et al., 2005a, 2007), which is charged by thunderstorms to a potential of several hundred thousand volts (Roble and Tzur, 1986) and drives a vertical current through the fair weather atmosphere's columnar resistance (~100 ohm). The current causes weak electrification of stratified clouds (Harrison and Carslaw, 2003) and produces a vertical potential gradient in the atmosphere near the Earth's surface. The circuit is closed by a horizontal current flowing through the highly conducting Earth and the ionosphere, and by vertical currents from the ground into the thunderstorm and from the top of the thunderstorm to the ionosphere. Figure 2 shows a schematic representation of a section of the global circuit through the dawn-dusk magnetic meridian (Tinsley, 2008), with the land/ocean surface forming one spherical highly conducting circuit element and, above that, the ionosphere forming the concentric outer shell of the circuit. In this model the upper conducting boundary is taken to be at about 60 km altitude. Additional generators are shown near the poles due to the solar wind moving past the magnetosphere of the Earth and producing a **V × B** electric field.

In the circuit T and S represent the column resistances of the Troposphere and the Stratosphere. The column resistance at a location varies due variations in surface altitude (orography), variations of aerosol concentrations and ion production by galactic cosmic rays (GCR) and other space particle fluxes, etc. To account for variations with latitude a subscript on T and S is used for equatorial, low, middle, high and polar latitudes. Even very small changes (1~ 3%) in the cosmic ray flux in the equatorial region due to variations in the solar wind may affect the thunderstorm charging current and the ionospheric potential. The variation in GCR flux could produce a 5% variation in T and S at low latitudes and, at high latitudes, ~10 – 20% (Tinsley, 2008). In addition to GCR, solar energetic particle (SEP) events could also change the resistance of the atmosphere in the polar cap region (Kokorowski et al., 2006). Variations in the column resistance of the troposphere and stratosphere cause variations in the vertical current $J_z$. Apart from external factors, $J_z$ also changes due to internal forcing. For example day-to-day variations in ionospheric potential due to changes in highly electrified deep convective clouds, mainly in the humid low latitude land areas of Africa, the Americas and northern Australia/Indonesia, produce changes in $J_z$ (Tinsley et al., 2007).

The relaxation time in the global atmospheric electric circuit varies with altitude; it is ~ $10^{-4}$ sec at 70 km altitude, increasing with the decrease in altitude to about 4 sec near 18 km, and 5-40 min near the Earth's surface. Measurements have never shown a complete absence of the fair-weather electric field, suggesting continuous operation of thunderstorms and other generators in maintaining the current flow in the GEC. Considering the charge carried to the upper atmosphere to be ~ $2 \times 10^5$ C, the potential drop between the ionosphere and ground is ~ 250 kV (Roble, 1985; Rycroft et al., 2000; Singh et al., 2004a). The traditional role of thunderstorms is to charge the ionosphere and to maintain the Earth-ionosphere potential difference, whereas TLE phenomena may discharge and reduce the potential difference, with each jet removing 30 C from the ionosphere (Hu et al., 2002; Sato et al., 2002). The charge removed by each jet thus accounts for ~ 0.015% of the total charge. The contribution of sprites to the global DC atmospheric electric field is of the order of 44 mV/m (Fullekrug, 2004); this is much smaller than the sensitivity of the electric field mills which measure the fair weather potential gradient ($\geq$ 1V/m). Fullekrug and Rycroft (2006) proposed a coupled model for the global static and dynamic electric fields derived from Maxwell's equations and showed that this small contribution can be measured with conventional radio equipment at frequencies ≤ 4 Hz.

The GEC provides a good framework for understanding solar-terrestrial-weather relations (Markson, 1978). These relate solar sector boundary crossings in interplanetary space to increasing lightning frequency (Reiter, 1972), thunderstorm activity (Cobb, 1967; Reiter, 1972) and vorticity area index (Markson, 1978). Cloud-to-ionosphere discharges affect the GEC by producing transient plasma in the mesosphere, which causes an enhancement of electrical conductivity for a short duration. Further, observations indicate that the ELF waves associated with TLEs reduce the electrical potential between the ionosphere and cloud (Rycroft, 2006). Thus it becomes necessary to modify the conventional GEC picture by including the contributions of gigantic jets, blue jets, elves and sprites to changes of electrical conductivity and ionospheric potential. The detailed knowledge of characteristic properties of these emissions along with frequency of events will help us to understand their contribution to the GEC (Pasko, 2003). Recent developments in this area and possible linkages with several other phenomena such as cosmic rays, atmospheric aerosols, weather and climate, sprites, blue jets and elves, etc. are discussed by Siingh et al. (2007).

**2.2.2 Lightning and Climate:**

Measurements show variations of the atmospheric electrical conductivity ($\sigma$) due to variations in radon, local aerosols and humidity exacerbated by vertical convection and turbulence. Such vertical current $J_z$ changes could dominate short term variations. Burns et al. (2007) showed high latitude surface pressure changes in response to $J_z$ changes on a day-to-day basis. They considered $J_z$ changes caused by ionospheric potential changes due to variations in the low latitude highly electrified convective cloud generators of the global circuit. Tinsley et al. (2007) showed that both on day-to-day and millennial time scales such a response to solar activity can be understood in terms of cloud microphysical responses to the $J_z$ changes. The flow of $J_z$ through gradients in conductivity at the cloud boundaries create space charges which rapidly attach to droplets and aerosol particles including cloud condensation nuclei (CCN) and ice forming nuclei (Tinsley et al., 2001). This increases the production rate of primary ice by contact ice nucleation, leading to an enhanced rate of precipitation, which may affect storm dynamics and the general circulation (Tinsley and Deen, 1991). On the other hand the enhanced scavenging of larger cloud condensation nuclei (CCN), along with other aerosol particles, may protect the smallest CCN and other small particles from scavenging by other processes such as Brownian diffusion and phonetic scavenging (Tinsley, 2004). This causes an enhancement in CCN concentration and a narrowing in droplet size distribution during cloud formation. As a result precipitation may reduce and cloud lifetime may increase.

Droplets with positive charge in the range 80 – 90 e were measured in downdrafts for droplets of radii 6 – 8 μm at cloud tops, whereas near the cloud base droplets were found with negative charge in the range 50 – 70e, in updrafts (Beard et al., 2004). The magnitude and sign difference of the charges near the two boundaries are consistent with the calculations of droplet charging resulting from the flow of $J_z$ through clouds (Zhou and Tinsley, 2007; Tinsley et al., 2007). The time constants for aerosol and droplet charging with ambient ions could range from minutes to hours which are comparable to typical convection and turbulence characteristic times (Tinsley, 2008). This shows that time dependent cloud and aerosol charging models including turbulence and convection are needed.

The variation in $J_z$ and hence variation in GEC is directly related to microscopic and macroscopic processes associated with clouds/lightning and would produce an integrated effect on climate that could dominate over short term weather and climate variations (Williams, 2005). The small variations in electric field near a cloud surface affects cloud formation and hence the temperature of the atmosphere. Williams (1992) reported an extremely non-linear increase in tropical lightning rate when temperature rose above a critical threshold (nonlinear sensitivity of thunderstorm activity to temperature). He also showed a high correlation between monthly mean tropical surface air temperatures and SR measurements of global lightning activity. Markson and Price (1999) reported a positive correlation between ionospheric potential and global temperature, ionospheric potential and global lightning/deep cloud index, global lightning and global temperature. They suggested that warmer conditions lead to more deep convection resulting in a higher ionospheric potential. Price (1993) showed good agreement between the diurnal surface temperature changes and the diurnal variability of GEC. He suggested that a 1 % increase in global surface temperature could result in a 20 % increase in ionospheric potential. Fullekrug and Fraser-Smith (1997) have inferred global lightning and climate variability from the ELF magnetic field variations. In the global framework the response of lightning and electrified clouds to temperature and changes in temperature have been analyzed and many time scales, including semi-annual and annual have been reported (Williams, 2005). In the case of the semi-annual variation, a $1^0$C increase in temperature above the threshold value may result in a 50% increase in global lightning frequency (Williams, 2005).

A close relationship has been shown between (i) tropical surface temperature and monthly variability of SR (Williams, 1992), (ii) ELF observations in Antarctic/Greenland and global surface temperature (Fullekrug and Fraser-Smith, 1997), (iii) diurnal surface temperature changes and the diurnal variability of the GEC (Price, 1993) and (iv) ionospheric potential and global/tropical surface temperature (Mulheisen, 1977; Markson, 1986; Markson and Price, 1999). Reeve and Toumi (1999), using satellite data, showed agreement between global temperature and global lightning activity. Price (2000) extended this study and showed a close link between the variability of upper troposphere water vapor (UTWV) and the variability of global lightning activity. SR measurements could reveal excellent agreement between the variation of UTWV with surface temperature and lightning activity. UTWV is closely linked to the other phenomena, such as tropical cirrus cloud, stratospheric water vapor content, and tropospheric chemistry (Price, 2000). Recently, Williams et al. (2005) discussed in detail the physical mechanisms and hypotheses linking temperature and thermodynamics with lightning and the global circuit. This clearly shows that the study of physical processes involved in the global electric circuit, the variability of global lightning activity and its relation to surface temperatures, tropical deep convection, rainfall, upper tropospheric water vapor content, and other important parameters that affect the global circuit are essential to understand the global climate system, which is so essential for the betterment of human society.

**2.2.3 Lightning and precipitation**

Cloud-to-ground lightning discharges have been used to estimate rainfall. Zipser (1994) used the ratio of monthly rainfall to the number of thunderstorm days to study the rainfall and thunderstorm relation for West Africa. Petersen and Rutledge (1998) used the total rain mass and CG flash density to examine the relationship over large spatial and temporal scales for several different parts of the globe, and found them to vary significantly, depending on air-mass characteristics and cloud microphysics. Earlier studies showed some positive relationship between lightning and area averaged convective rainfall (Marshall and Radhakant, 1978, Piepgrass and Krider, 1982, Tapia et al. 1998), and total lightning flash rate and convective rainfall (Goodman and Buechler, 1990, Chez and Sauvageot, 1997, Peterson and Rutledge, 1998). The relation between rainfall and lightning is generally expressed in terms of the Rainfall-Lightning ratio (RLR). This ratio estimates the convective rainfall volume per cloud-to-ground lightning (CG) flash. The RLR depends on thermal and microphysical characteristics of the thunderstorm, its location, local climatology and convective regime (Williams et al., 1989, Tapia et al. 1998, Seity et al. 2001, Lang and Rultedge, 2002) and varies over a wide range. The nature of the relationship is less certain over the oceans and, in particular, over the tropical oceans (Petersen and Rutledge, 1998; Zipser, 1994). Rakov and Uman (2003) and Kempf and Krider (2003) have shown that the RLR varies from $2 \times 10^4$ m$^3$ per CG flash to $2 \times 10^7$ m$^3$ per CG flash, depending upon geographical location and season of the year. It has been also observed that severe storms produce lower RLR values than ordinary thunderstorms.

Buechler and Goodman (1990) and Williams (1992) have shown inverse correlations between RLR and Convectively Available Potential Energy (CAPE). Considering wet bulb temperature ($T_W$) to be a reliable indicator of CAPE in the tropics Williams and Renno (1993) and Williams (2005) showed that the lightning flash rate has a good correlation with CAPE. Williams (1992) has also shown that the ratio of total rainfall to observed lightning in convective rain can be negatively correlated to the Convective Available Potential Energy (CAPE) of that thunderstorm. These discussions clearly show that additional measurements are required before any firm conclusion can be drawn.

In the modeling of weather forecasting, knowledge of the latent heat rate is required. This can be derived from lightning data and we can improve upon it if we have a good amount of data showing lightning-rainfall relationships. For example, Alexander et al. (1999) reported a relatively good correlation between convective rainfall and lightning rates during a large storm in 1993, and showed improved numerical forecasts by assimilating latent heating rates derived from lightning data. On the basis of the comparison made between lightning rates measured by Pacnet and convective rainfall data obtained from TRMM (Tropical Rainfall Measuring Mission) microwave sensors for a variety of storm systems over the central North Pacific, Pessi et al. (2004) suggested that the lightning data over the Pacific can be assimilated into numerical models as a proxy for latent heat release in deep convective clouds.

**2.2.4 Lightning, Atmospheric Aerosols and Cloud Processes**

Aerosol particles are generated by gas-to-particle (GPC) or drop-to-particle conversion processes. Trace gases also play a significant role through the aqueous-phase chemistry of the atmosphere and contribute to the generation and destruction of aerosols. Aerosol particles in the atmosphere partly act as nucleation centers for cloud formation and partly as ice forming nuclei. The number of aerosol particles which can serve as cloud condensation nuclei (CCN) increases with increasing supersaturation because even smaller particles participate in nucleation at higher supersaturation values. Observations show that continental air masses are generally richer in CCN than maritime air masses (Twomey and Wojciechowski, 1969). The Aitken nuclei produced by gas-to-particle reactions under supersaturation conditions are likely to become cloud nuclei (Vohra et al., 1970; Vohra and Nair, 1970). Hegg et al. (1980) observed a shift of the particle size spectrum towards larger sizes on passing from the upstream to downstream side of a cloud. A numerical simulation of the growth and subsequent evaporation of a convective cloud produced a similar result. Heintzenberg et al. (1989) observed a pronounced shift of the size distribution to larger sizes due to the processing of clear air particles by cloud. Hoppel et al. (1994) proposed that drop-to-particle conversion is the cause for the double maxima found in maritime aerosol particle spectra.

Aerosol particles and trace gases differing vastly in their physical and chemical characteristics are removed from the atmosphere through the processes of nucleation scavenging (incorporated in cloud drops, raindrops or ice crystals), impact scavenging (collected by collisions with cloud drops, raindrops or ice crystals) and gas scavenging (absorption of trace gases by cloud particles and raindrops). As a result of the various scavenging processes, the gas-phase and aqueous-phase chemistry of the atmosphere becomes complex. The whole variety of anthropogenic aerosol particles and trace gases injected into the atmosphere makes the situation still more complex.

It is difficult to distinguish the effect of aerosols from that of thermodynamics/dynamics on thunderstorm activity. Williams et al. (2002) conducted experiments to verify the enhanced lightning activity predicted by the aerosol hypothesis and arrived at different results during the lightning-active pre-monsoon and during the less active wet season in Brazil. Williams and Stanfill (2002) supported the thermal hypothesis to explain the lightning activity variation with island area acting in oceans as heat islands. Lyons et al. (1998) and Murray et al. (2000) attributed the enhancement in the positive cloud-to-ground flashes in North America to incursion of smoke from fires in Mexico and subsequent ingestion by these thunderstorms. However, the effect of smoke from biomass burning on thunderstorm activity observed in Brazil does not support the above result (Williams et al., 2002). MacGorman and Burgess (1994) reported clustering of positive ground flashes below storms which developed in a strong instability. In fact the possible role of the model results of Baker et al. (1999) indicate that the lightning flash rate is proportional to the fourth power of vertical velocity of aerosols whereas field experiments undermine this sensitive relationship (Williams 1992; Rutledge et al., 1992). This point should be investigated in detail. Further studies are required to explore the relationships between Convective Available Potential Energy (CAPE) in dry and moist convections and the potential temperature ($\theta$ or $\theta_w$) in the tropics and the role of cloud base height in transferring CAPE to updraft kinetic energy in thunderstorms (Williams and Stanfill, 2002; Williams and Renno, 1993; Lucas et al., 1994; Williams et al., 2002).

**2.2.5 Radio wave propagation and thunderstorms**

Radio communications in the VHF and UHF bands have seen a rapid growth during the last two or three decades. However, the quality of electromagnetic waves in these frequency bands undergoes deterioration in the presence of water vapor, clouds and hydrometeors such as rain, hail and fog which absorb, attenuate and scatter these waves. The water molecule has a permanent electric dipole moment and, being an asymmetric top molecule, exhibits a complex absorption spectrum of rotational transitions with a peak at 22.23 GHz and strong lines at 183 GHz and 325 GHz.

During rainy periods, the water droplets intercepting a radio path act as an imperfect conductor. Due to their large dielectric constant (81 times that of air) water droplets attenuate these waves. In addition, these displacement currents induced in the droplet of rain or fog act as sources of scatter for incident radiation. Non-spherical raindrops cause depolarization. The role of thunderstorms on the propagation of these high frequency waves should be studied both experimentally and through modeling work.

## 3. Transient Luminous Events above Thunderstorms

Thunderstorm charges and electric fields build up with time until a breakdown threshold is reached. The charge structure could be a complex structure leading to upward and downward lightning discharges, which are associated with Transient Luminous Events occurring at high altitudes in the Earth's atmosphere above thunderstorms. TLEs include sprites, elves, jets, gigantic jets, blue starters, etc. (Neubert et al., 2008 and references therein). The possibility of discharges above a thundercloud at high altitudes was predicted by Wilson (1925) based on electric field computations due to the residual charges of the cloud just after a positive cloud-to-ground discharge.

The observations of spectacular optical flashes by Franz et al. (1990) initiated experimental and theoretical studies in this area. Ground and aircraft campaign were conducted in the USA, Australia, Japan, Taiwan and Europe, and a large number of events concerning TLEs have been documented and studied (Rodger, 1999; Chern et al., 2003; Fullekung et al., 2006; Pasko, 2007; Neubert et al., 2005, 2008, Arnone et al., 2008). Amongst TLEs, sprites are widely observed/studied phenomena and they are usually found over Mesoscale Convective Systems (MCS), and not over ordinary isolated thunderclouds (Sentman et al., 1995; Lyons, 1994; Boccippio et al., 1995; Lyons, 1996; Lyons et al., 2008). MCS are characterized by laterally extensive regions of stratiform precipitation with a total area more than an order of magnitude greater than the area of an ordinary thunderstorm. Sprites are induced by +CG lightning strokes possessing large charge moment changes. Blue jets are slow moving "fountain of blue light" from the top of the cloud, whereas elves are lightning induced rings of light that can spread over ~ 300 km laterally at around 90 km altitude in the lower ionosphere. Su et al. (2003) have reported six gigantic optical jets from oceanic thunderstorms that establish a direct link between a thundercloud (~ 16 km altitude) and the ionosphere at 90 km. ELF waves were detected in only four of these events and no cloud-to-ground lightning was observed to trigger these events.

**3.1 Morphology and discharge mechanisms**

Sprites may occur in clusters of two, three or more "carrot" shaped emissions of ~ 1 km thickness over a horizontal distance of 50-100 km, with the separation between sprite elements of ~ 10 km (Neubert et al., 2005). Sprites may also occur as single luminous columns termed C-sprites (Wescott et al., 1998). Telescopic imaging revealed sprites with intertwined discharge channels and beads to scales down to 80 m diameter and smaller (Gerken et al., 2000; Mende et al., 2002). The optical intensity of a sprite cluster as observed by TV-frame rate cameras is comparable to that of a moderately bright auroral arc (Sentman et al., 1995). The brighter region is in the altitude range 65-85 km with most of the intensity in the red, and with wispy faint blue tendrils extending down to 40 km or at times as low as the cloud top (Wescott et al., 2001). High speed photometry shows that the duration of sprites is from a few ms to ~ 200 ms (Winckler et al., 1996; Armstrong et al., 1998). High speed imaging reveals that the discharges are initiated at ~ 65 km altitude; they propagate downwards and shortly after upwards (Stanley et al., 1999; Moudry et al., 2003). Recent imaging at 10 kHz frame rates shows the formation and propagation of streamers and resolves the streamer heads (McHarg et al., 2007). Stenback-Nielsen et al. (2007) showed that sprite emission rate on the shortest time scale can reach 1 - 500 GR.

Based on ELF/VLF observations, Ohkubo et al. (2005) suggested that IC discharges play a significant role in sprite generation. The IC flash and sferic activity is caused by breakdown processes inside the clouds feeding the continuing currents of +CG discharges. Using the Euro-sprite data, this point has been examined by Neubert et al. (2008), who showed that column sprites are generally characterized by a short time delay relative to the causative + CG discharge with high peak current, little IC activity and short, intense bursts of broadband VLF radio wave activity, whereas carrot sprites are associated with longer time delays, large IC activity and weaker, longer lasting bursts of lightning discharges. This shows that IC discharges play an important role in the generation of carrot sprites but may not play significant role for the impulsive column sprites. Neubert et al. (2008) have presented a sprite which is laterally displaced from the +CG and have proposed a process in which a +CG discharge may transfer charge from a remote region of the storm to the ground. A laterally displaced sprite was also reported by Mazur et al. (1998).

The above features of sprites depend on the thunderstorm source fields (Pasko et al., 1997; Valdivia et al., 1998) and the very rapidly varying field composed of the directly emitted electromagnetic field from the discharges and components reflected from the ground and the ionosphere (Cho and Rycroft, 2001). Local perturbations of the mesosphere may affect sprite morphology. For example, neutral gas density perturbations caused by gravity waves generated in the underlying thunderstorm could modulate the local threshold electric field for the sprite discharges (Sentman et al., 2003), with lower densities requiring a lower threshold electric field. An example of a sprite cluster that could have been formed by such a process has been reported by Neubert et al. (2005, 2008). The small scale (~ 1 km) structure in the sprite could also be produced by the perturbations of electric conductivity induced by meteor trails (Suszcynsky et al., 1999; Symbalisty et al., 2000). This shows that there is a complex link between medium and small scale structures of sprites, gravity waves and medium and small scale structures of the mesosphere.

Sprites are produced due to the excitation of atmospheric constituents by collisions with free electrons accelerated to sufficient energy. Early observations suggested a strong signature of $N_2$ (1 P) emissions and an absence of the $N_2^+$ Meinel emissions (Swenson and Rairden, 1998). Emissions in the Meinel band are strongly quenched by the atmosphere and hence they are not observed (Armstrong et al., 1998). The second proposed mechanism involves the ionization of neutral constituents in air breakdown cascades by the polarization electric field over the cloud tops following a + CG discharge (Pasko et al., 1995, 1996, 1997). Breakdown occurs when streamers are formed with high space charge density in the streamer tips. Model calculations and observations compare well with energies of ~ 2 eV in the sprite proper (Morrill et al., 2002) and ~ 5-25 eV in the halo surrounding the sprite (Miyasato et al., 2003). This model is valid for both + CG and – CG lightning, although – CG lightning is less effective in sprite generation (Neubert et al., 2008). However – CG lightning has been observed to cause sprites on rare occasions (Taylor et al., 2008). Spectral observations from space show the existence of significant impact ionization of $N_2$ (Mende et al., 2005; Frey et al., 2005), which has also been inferred from ELF observations of currents associated with sprites (Cummer and Inan, 1997; Cummer, 2003). The third process involves electrons with energy > 5 keV, generated during cosmic ray ionization of atmosphere (Gurevich et al., 1992, 1999; Roussel - Dupre et al., 1994). This mechanism is also called a runaway electron discharge process. In this process the required threshold field is less by a factor of 10 than the classical discharge process. Hence the runaway discharge process seems to preclude classical air breakdown (Roussel - Dupre and Gurevich, 1996; Taranenko and Roussel - Dupre, 1996).

In order to explain optical emissions above thunderstorms, the space and time evolutions of lightning generated electromagnetic pulses were studied using two dimensional numerical simulations (Rowland et al., 1995; Inan et al., 1996; Veronis et al., 1999). Cho and Rycroft (1998), using electrostatic and electromagnetic codes, simulated the electric field structure from the cloud top to the ionosphere and tried to explain the observation of a single red sprite. To explain clusters of sprites, they suggested that the positive charges are distributed in spots and so may lead to clusters of sprites. The numerical simulations are based on the finite difference time domain treatment of Maxwell's equations. The redistribution of charge and the electromagnetic pulse from the lightning discharge may accelerate electrons, heat and ionize the atmosphere, which will result in nonlinear phenomena such as runaway breakdown (Rycroft and Cho, 1998; Rowland, 1998). The lightning generated electromagnetic pulse modifies the electron density and collision frequency of the ionosphere which changes the electrical conductivity of the atmosphere (Holzworth and Hu, 1995); this change may be used to explain the generation of elves (Nagano et al., 2003).

Observations support the idea that all the processes discussed above could play role in the generation of sprites and other luminous events above thunderstorm. Neubert et al. (2008) have discussed the merits and limitations of these processes and the additional observations needed to ascertain efficacy of these processes.

**3.2 Effects of TLEs on Mesosphere (ionization, infrasound, $NO_x$ production, and atmospheric dynamics):**

Transient luminous events (sprites, elves, jets, etc.) perturb the upper atmosphere by changing its electrical properties, perturbing atmospheric constituents and altering atmospheric dynamics. The perturbation in electrical conductivity of the mesosphere can be evaluated by measuring changes in the amplitude of VLF waves from transmitters and propagating in the Earth-ionosphere waveguide. Amplitude perturbations in 90 % cases were observed when sprites were generated by +CG discharges. In the case of + CG and – CG discharges which were not associated with sprites, an amplitude perturbation was observed only in a few cases. The sprite related VLF waves had long recovery times (~ 30 - 300 s), which suggests spatially extended and diffuse regions of electron density increases at altitudes higher than 75 km. This agrees with the theoretical prediction of air breakdown in the upper D-region during sprite occurrence triggered by strong quasi-static electric fields (Mika et al., 2005). Mika (2007) has discussed experimental data in which the incident VLF transmitter signal seems to be scattered from horizontally extended diffuse regions of electron density enhancements, most likely associated with halos or diffuse regions of the upper part of carrot sprites, rather than small scale streamers observed at lower altitudes.

The electromagnetic pulse (EMP) generating elves also create ionization (Taranenko et al., 1993; Rowland, 1998), which depends on its intensity. The EMP may be sufficient to cause elves and ionization at ~90 km altitude but not ionization at lower altitudes which affects VLF wave propagation.

Liszka (2004) suggested the generation of infrasound waves by sprites, whose signatures were detected by a network sensors in Sweden (Liszka and Hobara, 2006). The shape of the chirp signature in the spectrograms of infrasound can be explained by the horizontal size of the sprite (Farges et al., 2005). A theoretical model based on heating a vertical region in the mesosphere has been proposed to explain the amplitude of infrasound from a sprite (Pasko et al., 1998; Farges et al., 2005; Pasko and Snively, 2007). Measurements of the rotational intensity distribution of $N_2$ molecular bands may be potentially used for remote sensing of variations of gas temperature in sprite discharges (Pasko, 2007). Neubert et al. (2008) have concluded that sprite detection by infrared is an attractive alternative to optical detection, because it is not limited by clear viewing condition or by the absence of daylight. The infrasound signature makes automatic detection of sprites possible (Ignaccolo et al., 2008).

Pasko (2006) summarized the optical emissions associated with sprites, which include the first positive ($1PN_2$) and second positive ($2PN_2$) band systems of $N_2$, Lyman-Birge-Hopfield band system of $N_2$ ($LBHN_2$) and the first negative band system of $N_2^+$ ($1N\ N_2^+$), which have excitation energy thresholds in the range ~ 7.35 – 18.8 eV and lifetimes at 70 km altitude in the range ~ 69 ns – 14 µs. Green et al. (1996) using energy dependent electron excitation cross sections and laboratory data analyzed the spectrally resolved emission obtained by Mende et al. (1995) and Hampton et al. (1996) and concluded that the sprite electrons have sufficient energy to dissociate and ionize $N_2$. They also estimated the electric field driving sprite phenomenon to be 100 – 200 $Vm^{-1}$ at 70 km altitude.

The strong blue emissions associated with $1NN_2^+$ and $2PN_2$ band systems originating in the streamer heads are expected to be produced during the early sprite development period. This is in agreement with narrow band photometric and blue light video observations of sprites (Armstrong et al., 1998; 2000; Suszcynsky et al., 1998; Morrill et al., 2002) indicating short duration bursts of blue optical emissions appearing at the initial stage of sprite formation. The time averaged optical emissions are dominated by red emissions associated with the $1PN_2$ band having the lowest energy excitation threshold (~ 7.35 eV) which can be produced by relatively low electric fields in the streamer channels. This is in agreement with reported sprite observations (Mende et al., 1995; Bucsela et al. 2003; Morrill et al., 1998; 2002; Takahashi et al., 2000). Pasko (2007) has documented important similarities between optical emissions associated with streamers in sprite discharges and emissions from pulsed corona discharges in laboratory experiments and emphasized need for further studies of processes related to the vibrational excitation of the ground state of $N_2$ molecules to understand emissions originating from the $B^3\Pi_g$ and $C^3\Pi_g$ states of $N_2$ and NO γ-bands emissions, during both initial and post discharge stages of a sprite discharge.

The sprites provide a link between tropospheric processes in the thunderstorms and mesospheric processes in the upper atmosphere. Hiraki et al. (2002) suggested that sprites may change the concentration of $NO_x$ and $HO_x$ in the mesosphere and lower atmosphere. These chemical changes have impact on the global cooling or heating in the middle atmosphere (Galloway et al., 2004; Singh et al., 2005; Schumann and Huntrieser, 2007). Nitrogen oxides are critical components of the troposphere which directly affect the abundance of ozone and hydroxyl radicals (Crutzen, 1974). Ozone absorbs solar ultraviolet radiation and controls the dynamic balance of the atmosphere. $NO_x$ creates ozone in the troposphere and destroys it in the stratosphere and mesosphere. The concentration of $NO_x$ in the mesosphere is enhanced by transient events such as auroras and solar proton events (Crutzen and Solomon, 1980). The vertical transport of $NO_x$ by the neutral wind is an important process to control $NO_x$ concentration in the thermosphere (Saetre et al., 2007). TLEs also affect the concentration of $NO_x$ in the stratosphere and mesosphere and the processes involved could be similar to those in the aurora. The production of $NO_x$ is through energetic neutral atom conversion where the heating of the atmosphere by the continuing current in the ion channel of the lightning stroke allows suprathermal oxygen atoms to react directly with $N_2$ (Balakrishnan and Dalgarno, 2003).

The effects of TLEs on the chemistry of the middle atmosphere have been studied using theoretical models and simulation codes. Enell et al. (2008) have used a particle code based on independent streamer propagation models whereas Sentman et al. (2008) used a fluid code for streamer propagation. Gordillio-Vazquez (2008) used a plasma code with a Boltzmann solver. In the computation it is assumed that suprathermal atoms do not occur in a sprite, although they may occur in the longer-lived blue jets and gigantic jets (Su et al., 2003). Enell et al. (2008) estimated that the total production of $NO_x$ is around 5 times the background in the streamers, whereas the model of Gordillio-Vazquez (2008) predicts NO and $NO_2$ enhancements of a factor of 10. Neubert et al. (2008) estimated the total number of NO molecules produced in a streamer ~ $1.5 \times 10^{19}$, with an estimated average total production on the order of $10^{23} - 10^{25}$ molecules per event. The estimate is based on the dimensions of a typical sprite event (Gerken et al., 2000). Sentman et al. (2008) estimated a production on the order of $5 \times 10^{19}$ molecules per streamer, which is consistent with the above results. Neubert et al. (2008) have shown that considering the global occurrence rate of sprites to be 3/minute, the total global production of $NO_x \sim 10^{31}$ molecules per year, which is of the same order as the minimum production of $NO_x$, $N_2O$, $N_2O_5$ and $HNO_3$ by solar proton events during a quiet year. The stratospheric production by oxidation of $N_2O$ is $\sim 10^{34}$ molecules per year. This shows that the production of $NO_x$ by sprites on the global scale is quite small. However, during an intense thunderstorm, there can be a significant impact on the local production and budget of $NO_x$. Enell et al. (2008) discussed that the NO enhancement (2 – 10%) in the 50 – 60 km altitude range could decrease the ozone concentration by a few percent. However, significant ozone perturbations by sprites are unlikely. Measurements from the GOMOS instrument on ENVISAT showed enhancements on local basis but these were not found on the larger regional scale (Rodger et al., 2008).

The perturbations in ozone affect the atmospheric dynamics in two ways, namely by perturbation in the absorption of solar radiation and secondly by affecting the local strength of the zonal winds which affects the propagation of planetary waves in the stratosphere. Sprites could affect the circulation in the middle atmosphere, which may also control the downward propagation of stratospheric temperature anomalies into the troposphere (Christiansen, 2001), which imply a possible impact of stratospheric ozone changes on the tropospheric climate. Several model computations have been made to assess the strength of ozone perturbation needed to produce a significant change in the stratospheric dynamics and the extent to which this change propagates down to the troposphere (Christiansen et al., 1997; Berg et al., 2007). The computations show that the response of ozone perturbations on the middle atmosphere dynamics is nonlinear (Neubert et al., 2008).

Pasko et al. (2002) have reported a video recording of a blue jet propagating upwards from a small thundercloud cell to an altitude of about 70 km. As relatively small thundercloud cells are very common in the tropics, it is probable that optical phenomena from the tops of the clouds may constitute an important component of the GEC. Because optical phenomena occur in the upward branch of the global electric circuit above the thunderstorms they are likely to influence only the upper atmosphere conductivity. Moreover, since they occur much less frequently (only one sprite out of 200 lightning discharges), because of their association with intense lightning discharges (Rycroft et al., 2000; Siingh et al., 2005a; Siingh et al., 2007), they may not play a major role in the GEC (Rycroft et al., 2000). Further research activity in this area is required.

## 4. Electromagnetic radiation from lightning and TLEs

Lightning and TLEs are the results of discharges between the charges inside the cloud and below the cloud on the surface of the Earth and above the cloud. According to classical electrodynamics, electric discharges radiate electromagnetic waves, whose frequency bands are determined by the time scales of different processes acting in the discharges. Cloud-to-ground lightning discharges radiate electromagnetic waves over a wide frequency range from a few Hz to several hundred MHz. However, the maximum energy is radiated in the VLF range, especially between 1 and 10 kHz. The intensities of electromagnetic waves generated during lightning discharges decrease with frequency f as $f^{-2}$ at lower frequencies and as $f^{-1}$ at higher frequencies. The dominant time scales of lightning discharge and sprite processes, which are responsible for the broadband electromagnetic emissions, vary from microseconds to some tens of ms (Fullekrug et al., 2006). The electromagnetic radiation from lightning discharges propagates to long distances and can be a powerful tool for studying the lightning, TLEs and the parameters of the medium through which it propagates. However, a sprite is triggered by a lightning discharge and occurs almost simultaneously with it; therefore it is difficult to separate the electromagnetic wave signature of a sprite from that of the causative lightning discharge.

**4.1 ULF Waves**

Ultra low frequency (ULF) waves (< 3 Hz) radiated by +CG lightning discharges and propagated to long distances have been recorded and analyzed. Even though the magnetometer could record traces of lightning discharges around the world (Fukunishi et al., 1997), this could be due to secondary/tertiary processes having much longer time scales which emit electromagnetic waves as well (Neubert et al., 2008). The general properties of ULF waves from the sprite associated lightning discharges were compared with those of a control group not associated with sprites, but a unique and identifiable difference could not be obtained (Bosinger et al., 2006; Neubert et al., 2008). All responses represent single, isolated, strong events called Q bursts (Ogawa et al., 1967), which exceed the average natural background noise level caused by world-wide thunderstorm activity at least by an order of magnitude.

**4.2 Schumann resonances**

Tropical lightning discharges excite the Earth-ionosphere cavity which resonates at the fundamental frequency and its harmonics, known as Schumann Resonances (SR), with the fundamental mode having an eigenfrequency of 8 Hz (Sentmann, 1995; Huang et al., 1999; Barr et. al., 2000; Singh et al., 2002). The amplitude of the Schumann modes is similar at all places. Variations in solar activity or nuclear explosions produce disturbances in the ionosphere and may also affect SR (Schlegel and Fullekug, 1999). Solar proton events cause an increase of frequency, Q-factor (i.e. reduced bandwidth of the resonance mode) and amplitude of the SR mode (Schlegel and Fullekrug, 1999). Sentman et al. (1996) examined SR measurements from California and Australia during the large solar storms in the fall of 1989 and found no measurable difference in SR intensities, although they found a sudden decrease in Q-factor of the second mode, which was attributed to small changes of middle atmospheric conductivities by energetic charged particles. The SR intensity depends upon the height of the ionosphere (Sentman and Fraser, 1999). It has solar cycle dependence and responds to solar flares, magnetic storms (Hale and Baginski, 1987) and solar proton events (Reid, 1986).

The principal features of SR are used to monitor global lightning activity (Heckman et al., 1998; Barr et al., 2000; Rycroft et al., 2000; Siingh et al., 2007), global variability of lightning activity (Satori, 1996; Nickolaenko et al., 1996) and sprite activity (Boccippio et al., 1995; Cummer et al., 1998a; Rycroft et al., 2000). Apart from locating the parent lightning flashes, Schumann resonance methods have been used to evaluate their vertical charge moments. Williams et al. (2007a) evaluated the charge moment threshold for a sprite located at a distance of ~ 16.6 Mm and showed that their results were consistent with similar measurements using identical methods made at a considerably closer distance (~ 2 Mm). The vertical charge moment change of the lightning is the key source parameter in the initiation of sprite (Huang et al., 1999; Hu et al., 2002). Williams et al. (2007b) have shown that the current moment spectra of sprite producing +CG lightning are redder than that of those which were not associated with sprite. This shows that SR can be successfully used to study the location as well as the properties of the sprite-associated lightning.

Recently Simoes et al. (2008) have used SR as a tool for exploring the atmospheric environment and the subsurface of the planets and their satellites. In fact the knowledge of the frequencies and attenuation rates of the principal eigenmodes provides unique information about the electrical properties of the cavity. They have developed models for selected inner planets, gaseous giants and their satellites and have reviewed the propagation process of SR waves in their atmospheric cavity so as to infer the subsurface properties.

Since thunderstorms are the main source of energy for SR and the GEC, their link with weather and climate should be developed (Williams, 1992; Price, 1993; Price and Rind, 1994). Such links could be in the form of electromagnetic, thermodynamic, climate and climate-change characteristics of the atmosphere. The physico-chemical processes involved in these phenomena should be studied both empirically and theoretically. If lightning is the main source for maintaining the ionospheric potential, the measurements of SR and ionospheric potential should produce identical results. The differences between the two results will indicate the contribution to ionospheric potential of other processes, such as corona discharge from elevated objects above the ground (Rycroft et al., 2007) or any other unknown process.

**4.3 ELF Waves**

Extremely low frequency (ELF, 3 Hz – 3 kHz) electromagnetic radiation in the frequency range 8 – 100 Hz propagates over long distances within the Earth-ionosphere cavity without significant attenuation. Hence observations of ELF waves can be used to monitor the global occurrence rate of lightning and sprites. ELF waves are generated by sprites during mesospheric breakdown (Fullekrug et al., 2001), with sprites appearing ~ 5 ms after the causative +CG lightning discharge having a strong continuing current. The generation of sprite-associated ELF waves is not clearly understood. Fullekrug (2006) developed a model of lightning electric fields above a thunderstorm to study the charge transfer within intense lightning. Based on the analysis of a large number of intense lightning discharges, Fullekrug et al. (2006) showed that current decays with an initial time constant of ~ 2 ms for 10 second and transfers ~ 60 C charge from cloud to the ground. Later on the time constant becomes ~ 40 ms and transfers the remaining charge (~ 170 C). The current associated with the total charge transfer heats the middle atmosphere.

The measurement of ELF waves at large distances from the source is used to derive information about sprites. The measured electromagnetic signal in a known frequency range is used to derive the source waveform current moment of the lightning discharge (Cummer, 2003), using a numerical propagation model (Cummer, 2000). The waveguide propagation effects and bandwidths of digital filters used in the evaluation of ELF spectrum limit the narrowness of the current pulse to be resolved. It is shown that sub-millisecond variations are not very important in the study of sprites; however, the short time scale information is relevant for elves and the sprite halo (Barrington-Leigh et al., 2001; Cummer, 2003).

A combination of high speed video images and ELF magnetic field measurements showed that most sprites were initiated when the lightning charge moment exceeded 300 – 1100 C km (Cummer and Stanley, 1999). Stanley et al. (2000) reported larger charge moment changes (~ 6000 C km) required to initiate day time sprites, which is expected because the higher daytime mesospheric conductivity inhibits the penetration of quasi-electrostatic fields to higher altitudes. Cummer (2003) presented a case where the charge moment charge was 120 C km and a weak sprite was observed. Thus the generation of the sprite and its delay time are dependent in a complex manner on the charge moment change, charge transfer rate, mesospheric conductivity and maybe on many other local parameters, which either may create substantial ionization in the mesosphere or may help in the mesospheric ionization process. For example, the gravity waves produced during thunderstorm convective motions may propagate up to the mesosphere, where they may decay and produce heating which may help in the ionization of mesospheric constituents. If this is true, then one may look for a possible relation between sprites and the temperature inversion at the mesosphere (Fadnavis et al., 2007). Observations also show that a significant fraction of sprites (~ 10%) contain a clear sprite current ELF signature, which is distributed evenly across lightning-sprite delays. This suggests that there may be some processes which are present in those sprites which contain strong sprite currents and that they may be absent in the remaining cases.

**4.4 VLF Waves**

Apart from ULF and ELF waves, lightning discharges are the source of VLF (3 – 30 kHz), LF (30 – 300 kHz), MF (0.3 – 3 MHz), HF (3 – 30 MHz) and UHF (> 30 MHz) electromagnetic waves. Whistlers were observed by Storey (1953) and Helliwell (1965). The collected vertical electric field data were analyzed to derive information about the lightning discharges (source properties), propagation features and parameters of the medium through which the waves have propagated (Storey, 1953; Helliwell, 1965; Singh et al., 1998a). Later on, the horizontal components of the wave fields were measured along with the vertical components. These measurements were used to determine the polarization and arrival direction of the waves (Sagredo and Bullough, 1973; Hayakawa et al., 1986) and used to pin-point where the downcoming whistler entered the Earth-ionosphere waveguide (Hayakawa, 1993).

In recent years during the Eurosprite 2003 campaign, the vertical electric fields across the frequency range from 3 kHz to 30 MHz were measured at a number of places in France (Neubert et al., 2008) and attempts were made to correlate the radio emissions with CG discharges and with optical observations of sprites. HF emissions were measured during the time interval when sprites were observed. The radiation from a lightning discharge and a sprite was separated, and they clearly showed that HF bursts were associated with sprites.

Whistler mode waves are right hand polarized with the upper frequency cutoff as either the local electron plasma frequency or gyrofrequency, whichever is less (Stix, 1992). Ionospheric plasma consists of electrons and different types of ions having different ion gyrofrequencies. Hydrogen ion whistlers, oxygen ion whistlers, and helium ion whistlers have been observed on satellites.

**4.4.1 VLF sferics**

Sferics (short for atmospherics, radio signals from lightning discharges) propagate in the Earth-ionosphere waveguide with low attenuation, typically ~ 2-3 dB/1000 km (Davies, 1990). The waveform, spectrogram and power spectrum of a typical sferic is shown in Figure 3 (a-c); this was recorded on 23 March 2008 at Allahabad, India. The spectrum shows greater dispersion at lower frequencies. The power spectra show that the major portion of the wave energy lies in the frequency range 5 – 15 kHz. The waveform is identical to that reported by Singh and Singh (2005a) using numerical simulation.

The dynamic spectra of sferics are explained by considering propagation though the Earth-ionosphere waveguide, where the D-region of the ionosphere acts as the upper boundary of the waveguide. Because of their reflection from the D-region sferics are widely used for D-region studies, which is one of the least studied regions because it is inaccessible by satellites (altitude range is too low) and by balloon-borne equipment (altitude range is too high). To obtain D-region electron density profiles rockets are used (Smith, 1969; Danilov and Vanina, 2001), but these have their own limitations as rockets cannot be launched frequently. Ground-based active measurements include HF-VHF incoherent scatter radars; however, these techniques are difficult to apply due to the low electron densities ($< 10^3$ el/cm$^3$) at night (Hargreaves, 1992). Cummer et al. (1998b) developed a technique, which is based on wideband, long distance VLF propagation effects observed in sferics. This technique measures average electron density profile across the entire path and is therefore capable of estimating the electron density of large region. Using the same technique, Cheng et al. (2006) derived the night-to-night variations of the D-region ionosphere electron density over the East coast of the United States and compared the measurements to the results of past nighttime rocket experiments made at similar latitudes. Some recent work with sferics has focused on understanding the delayed sferic component called a "tweek" (Outsu, 1960; Yamashita, 1978; Yano et al., 1989; Yedemsky et al., 1992). Ohya et al. (2003) estimated equivalent nighttime electron densities at reflection heights in the D-region ionosphere at low-middle latitudes by accurately reading the cut-off frequency of tweek atmospherics. They reported the equivalent electron densities ranged from 20-28 el/cm$^3$ at ionospheric reflection height of 80-85 km. Ohya et al. (2006) examined the response of the nighttime D-region ionosphere to the great magnetic storm of October 2-11, 2000, using an accurate analysis of the cut-off frequency of tweek atmospherics.

The morphological features of sferics have also been used to estimate the distance and geographic bearing of the source discharges and the ionospheric reflection height along the propagation path (Kumar et al., 1994; Hayakawa et al., 1994). Wood and Inan (2002), using VLF magnetic field measurements at Palmer, Antarctica, calculated the azimuths of sferics originating in North America (~ 10,000 km range). Furthermore, they compared arrival azimuth with flash level data from the National Lightning Detection Network (NLDN) and showed that 83.6% of the cases matched the reported NLDN flashes to within 2 degrees. Several ground-based networks have been developed to determine the time and location of individual lightning strokes and flashes accurately using the characteristics of sferics, such as Lightning-Mapping Arrays (Rison et al., 1999; Thomas et al., 2000), UK Met Office VLF system (Lee, 1986a, b, 1989), and Los Almos Sferic Array (LASA) (Smith et al., 2002).

### 4.4.2 Whistler Phenomena

Whistler phenomena discuss wave propagation and related properties of VLF waves having right hand polarization and wave frequency smaller than the electron gyrofrequency and electron plasma frequency. These waves have been received both on the Earth's surface as well as onboard satellites. The waves can propagate through the magnetospheric plasma in a ducted mode, a non-ducted mode and a prolongitudinal mode. Waves propagating in the non ducted mode can be received only onboard satellites.

#### 4.4.2.1 Ducted Mode of Propagation:

Whistler mode waves entering the ionosphere/magnetosphere with small wave-normal angles are trapped in a geomagnetic field aligned duct formed by the enhancement/depletion of electron density. Three dimensional ray tracing computations using realistic modes for the plasma density and geomagnetic field show that the waves with wave-normal angles less than $4^0$ could be trapped in the duct and received on the Earth's surface (Walker, 1976; Ohta et al., 1997). Strangeways (1999) showed a reduction in wave path excursions both in latitude and longitude as rays propagate upwards in a duct from low altitudes (~ 1000 km) to the equatorial plane. The ray tracing method is applicable when the duct radius is greater than the characteristic wave-length (Strangeways, 1999). Pasmanik and Trakhtengerts (2005) have studied whistler mode wave propagation in magnetospheric ducts of enhanced cold plasma density for the arbitrary ratio of the duct radius to the whistler wavelength (where the ray tracing method is not applicable). They have showed that in the case of the source exciting simultaneously several duct eigenmodes, the receiver should register the set of discrete signals each corresponding to one of the excited eigenmodes. The interval between the arrivals of the different eigenmodes depends on the duct parameters, especially on the width. The interval is longer for a narrow duct and smaller in the case of a wide duct, and the interval can vary up to some hundreds of milliseconds.

The above results have been used to explain the trace splitting reported by Hamar et al. (1992) for whistlers recorded at the ground station Halley, Antarctica, as well as on the satellite Intercosmos 24 (Lichtenberger et al., 1996). This result can also be used to explain the reported trace splitting in the whistler data recorded at Varanasi (Singh, 1999; Singh et al., 1999a) and at Jammu (Singh et al., 2004b). Recently, Streltsov et al. (2006) presented a numerical study of whistler propagation in the presence of plasma density gradients transverse to the geomagnetic field with parameters typical of the equatorial magnetosphere of the Earth. Their results showed that low frequency whistlers can be trapped and guided by these narrow channels both in the case of reduced density channels and enhanced density channels. They further showed that energy can leak from the channel in the case of high density ducting because it is impossible for the whistler inside the duct to couple to a propagating wave outside the channel.

When a lightning discharge is composed of multiple flashes and VLF energy propagates in the same duct, the series of whistlers having the same dispersion and separated in time are called "multiflash" whistlers. When a lightning discharge illuminates more than one duct simultaneously in the magnetosphere then "multi-path" whistlers are produced. The dynamic spectra of whistlers are controlled by the path-length, the distribution of electron density and the magnetic field along the path of propagation. In figure 4 we present dynamic spectra of four whistlers recorded at Allahabad (geomagnetic latitude = $16.49^0$ N, longitude = $155.34^0$ E, L = 1.09), which is a newly set up low latitude Indian station on 17 June, 2008, at 12:40 UT. The lower and the upper cut off frequencies are 2.28, 2.03, 1.68, 3.12 kHz and 7.12, 7.86, 5.63, 5.63 kHz, respectively. The dispersion of the whistlers is 18.26, 15.72, 16.41 and 17.25 $\sec^{1/2}$, respectively.

The characteristic difference between low and mid-high latitude whistler spectra is (a) the upper cut-off frequency of low latitude whistler is higher than those observed at mid-high latitudes, (b) the nose-frequency ( ~ 0.4 $f_{He}$, where $f_{He}$ is the equatorial electron gyrofrequency) for low latitude whistler is ~ 100 kHz or more and hence is not observed due to heavy attenuation (the absorption coefficient is minimum ~ 5 kHz and increases with frequency), and (c) the dispersion is smaller than for mid-high latitude whistlers.

It is expected that a lightning discharge may illuminate a large area in the ionosphere through which wave energy may enter the magnetosphere to appear as a whistler wave in the conjugate region. Chum et al. (2006) have studied the penetration of lightning induced whistler waves through the ionosphere by investigating the correspondence between the whistlers observed on the DEMETER and MAGION-5 satellites and the lightning discharges detected by the European lightning detection network EUCLID. They demonstrated that the area in the ionosphere through which the electromagnetic energy induced by a lightning discharge enters the magnetosphere is up to several thousand kilometers wide. This suggested that stations situated within 1000 km could record the same whistlers. Using a statistical approach they assigned causative lightning discharges to the observed whistlers both on the Low Earth Orbit (LEO) satellites and to the whistlers observed at satellites orbiting at altitudes of ~ 5000 km. The results show that positive and negative lightning discharges have approximately the same efficiency in producing whistlers. However, no information could be obtained about the intra-cloud discharges, because the EUCLID network is rather insensitive to them. Comparing lightning data and one-hop whistlers obtained from an automatic detection system on the ground, Lichtenberger et al. (2005) and Collier et al. (2006) showed that the primary source of whistlers are cloud-to-cloud lightning and only a minor part of the whistler events are excited by cloud-to-ground discharges.

Recently, Ferencz et al. (2007) have presented dynamic spectra containing numerous fractional-hop whistlers recorded onboard DEMETER satellite and termed them "spiky whistlers". They appear to be composed of a conventional whistler combined with multimode sferics propagating in the Earth-ionosphere waveguide. To explain the dynamic spectra it is proposed that a part of energy from the cloud-to-ground lightning discharge propagating in multimodes in the Earth-ionosphere waveguide may leak from the waveguide and propagate upwards through the ionosphere/magnetosphere. These signals can be recorded on board DEMETER satellite. Ferencz et al. (2007) have presented a full wave numerical model to explain the observed dynamic spectra taking in to account the above concept.

Singh et al. (2008) recently presented unusual dynamic spectra of doublet and triplet whistlers recorded at the low latitude station Jammu during a daytime disturbed magnetic activity period. The analysis showed that these doublets and triplets belonged to mid latitudes having paths of propagation L = 2.63 and 2.67 for the whistler elements constituting doublets and L = 4.35, 4.39 and 4.43 for the triplet components. However, dispersion values for the doublets are 16.8 and 18.6 $\sec^{1/2}$, whereas for the triplets they are 81.9, 95.2 and 100.2 $\sec^{1/2}$, respectively. This shows that there is no correspondence between the L-value and dispersion of doublets. According to the dispersion of the doublets, the L-value is ~ 1.4, which is much less than the estimated value. This discrepancy may be due to the fact that the propagation path may not be along the geomagnetic field line. A similar discrepancy between L-value and dispersion has also been reported by Singh et al. (2004c, 2006a). In fact low dispersion may arise if the propagation path is small. In that case the total propagation path may lie in the ionosphere. In such a situation, the analysis method involving matched filtering technique may not be valid, because it requires a propagation path along the field lines and the electron density distribution is considered to be that for a diffusive equilibrium model. If the path lies in the upper ionosphere (L < 1.5) then plasma density irregularities and plasma blobs are frequently present there; these may affect wave propagation through the processes of scattering, diffraction, reflection and non-linear interaction (Sonwalker and Harikumar, 2000).

Singh et al. (2007a) analyzed whistlers recorded during the period January 1990 – December 1999 and showed the maximum monthly occurrence rate to be during January to March. The dependence of occurrence rate on geomagnetic disturbances was analyzed using the variation of $K_P$ index. They showed that the occurrence rate probability monotonically increases with $\Sigma K_P$ (daily sum) values. The occurrence rate is found to be greater than the average value for $\Sigma K_P \geq 20$. The results are in good agreement with the earlier studies from low latitude stations (Helliwell, 1965; Hayakawa, 1991; Singh, 1993). They have also shown that, as the intensity of a magnetic storm increases, the probability of whistler occurrence decreases. The enhancement in the whistler occurrence rate during geomagnetic disturbances at low latitudes is attributed to the formation of additional field aligned ducts (Somayajulu et al., 1972; Singh, 1993).

A marked seasonal variation is observed at every latitude. This is obviously due to the seasonal asymmetry of sources (lightning activity) in the conjugate hemisphere. The solar cycle variation exhibits a smaller occurrence rate during high solar activity and an enhanced one during low solar activity, in which D-region absorption might play a significant role. The latitudinal dependence of whistler occurrence rate shows a maximum around ~ $45^0$ geomagnetic latitudes (Helliwell, 1965). The lower latitude cut-off of whistlers is found to be ~$10^0$ geomagnetic latitude (Hayakawa et al., 1990). Below this latitude, the field aligned path is in the ionosphere where ionospheric plasma turbulence may affect field aligned wave propagation.

The analysis of whistlers recorded at a ground station yields information about the duct properties through which it has propagated. The analysis of occurrence rate and diffuseness of the whistler trace give an estimate of the lifetime and width of the duct. The lifetime of ducts may vary from a few minutes to many hours (Singh, 1993). However, statistical analysis suggests that the duct formation and decay are cyclic phenomena with the time scale of an hour (Hansen et al., 1983; Hayakawa et al., 1983; Singh and Singh, 1999). At low latitudes duct formation is difficult due to the required density enhancement (~ 100%), the presence of curvature in geomagnetic field, and plasma turbulence. In the absence of a duct, the observation of small dispersion low latitude whistlers are proposed to be due to propagation in the prolongitudinal (PL) mode with the angle between the wave normal and the geomagnetic field almost tending to zero (Singh, 1976; Singh et al., 1992; Singh, 1993). Recently, Kumar et al. (2007) reported an observation from the low latitude station Suva, Fiji, and interpreted the propagation mechanism to be PL mode, supported by the negative electron density gradient in the ionosphere that is enhanced during magnetic storms.

**4.4.2.2 Non-Ducted Mode of Propagation**

VLF waves entering the ionosphere/magnetosphere with large wave-normal angles to the geomagnetic field cannot be trapped in the field aligned ducts; they propagate in a non-ducted mode. Such waves suffer reflection at the point where the wave frequency equals the lower hybrid resonance frequency (Stix, 1992), and they do not reach the Earth's surface. They can only be received onboard satellites. Such whistlers are called as magnetospherically reflected (MR) whistlers and are composed of multiple discrete components (Smith and Angermi, 1968). Sometimes waves of higher frequency while propagating downward may not undergo lower hybrid resonance reflection; they propagate with their wave normal nearly perpendicular to the geomagnetic field but their ray direction remains nearly parallel to the Earth's magnetic field. Such a mode of propagation is called the proresonance (PR) mode, and the whistler traces are called walking trace whistlers (Walter and Angerami, 1969). When the wave frequency is lower than the lower hybrid resonance frequency and the wave normal angle is relatively small (but not small enough to be trapped in ducts) the waves propagate in the prolongitudinal mode (PL) and such whistlers can be received on low altitude satellites. They are called PL whistlers (Morgan, 1980) and they are considerably different from PR whistlers.

The other class of whistlers observed on satellites is proton/ion whistlers, propagating in the ion-cyclotron mode. The propagating frequency lies in the range between the proton gyrofreqency and the crossover frequency characterized by the frequency where the polarization becomes linear during the transition from right to left handed. Observed whistler traces are proton whistlers, helium whistlers, deuterium whistlers, etc. (Gurnett et al, 1965; Barrington et al., 1966). The crossover frequency yields information on ion composition (relative ion concentrations) (Gurnett and Shawhan, 1966; Shawhan and Gurnett, 1966).

### 4.4.2.3 Modeling whistler wave propagation

Whistler wave simulations for different plasma conditions have been carried out using Maxwell's equations, and the continuity and conservation of momentum equations (Ferencz, 1994; Ferencz et al., 2001; Singh et al., 2004c; Singh and Singh, 2005a,b,c). In these simulations, the lightning excitation source was represented by a Dirac delta function and the whistler's dynamic spectrum was obtained. Singh and Singh (2005a), using full wave analysis, derived equations for the whistler-mode signal propagating longitudinally through a one dimensional, inhomogeneous, weakly ionized magnetoplasma, in which wave-energy dispersion is caused by the interaction between electrons and wave fields, and dissipation of energy is caused by collisions between electrons and neutrals. The commonly known whistler spectrum is obtained from the space-time dependent wave field at a given point in space by the Fast Fourier Transform method. It is interesting to observe that the technique used under various conditions simultaneously explains low dispersion whistlers, nose whistlers, precursors (see later) and proton whistlers.

Singh and Singh (2005c) extended the results and included the effect of inter-particle collisions on the amplitude of the excited signals in different frequency ranges in terms of charge per unit length of the excitation source and the distance of propagation of the signal. It was shown that the amplitude of the signal depends upon the polarization mode, distance from the current source, the shape of the excitation current and frequency range considered. The technique has also been used to explain trans-ionospheric pulse pairs, TIPPs (Singh and Singh, 2005b). Singh et al. (2006b) used this technique to simulate the whistlers recorded at low latitudes and extended the frequency range to estimate the nose frequency, not experimentally observed on the dynamic spectra which are limited to <10 kHz. They showed that the method permits one to study nose frequency variations, which can be used to deduce electric fields in the magnetosphere (Singh et al., 1998 a,b).

Using two dimensional numerical simulations, the space and time evolution of the lightning generated pulse in the lower ionosphere was studied and the optical emissions generated were explained (Rowland et al., 1995; Inan et al., 1996; Veronis et al., 1999).

Cho and Rycroft (2001) developed a three dimensional code to calculate the optical emissions created by the electromagnetic pulse from a horizontal cloud-to-cloud discharge. These numerical simulations are based on the finite difference time domain treatment of Maxwell's equations. Nagano et al. (2003) using full wave analysis studied the ionospheric propagation of the lightning-generated electromagnetic pulse with a model including a horizontally stratified ionosphere, free space and the ground using Fourier spectral analysis and the plane wave expansion technique.

### 4.4.3 VLF Emissions

VLF emissions are whistler mode signals having complex structured and unstructured dynamic spectra observed both on the ground and onboard satellites. Unstructured emissions are characterized by a band limited spectrum for times ranging from a few milliseconds) to a few hours; they are called hiss emissions. Hiss emissions can be observed at repeated time intervals as pulsing hiss (Singh et al., 2008). Structured emissions exhibit coherent discrete frequency-time characteristics; these are emissions like chorus, periodic emissions, and quasi periodic emissions (Helliwell, 1965; Singh, 1993). VLF emissions may be generated following lightning discharges, or triggered by VLF waves of natural origin or by transmitted signals. Simultaneous measurements of VLF waves and charged particles onboard rockets and satellites suggest that the main source of energy for VLF emissions is the energetic electrons in the magnetosphere (Cornilleau-Wehrlin et al., 1985; Siingh et al., 2005a).

To pinpoint the generation mechanism of these emissions, it is essential to know the mode of propagation from the source region to the observation point, the source of energy and the mechanism which converts part of the energy into VLF emissions. Several theories have been proposed from time to time to explain the origin of these emissions. They differ significantly and can be classified into the following categories: Cerenkov radiation (incoherent and coherent), traveling wave tube mechanism, backward wave oscillator, cyclotron radiation and transverse resonance instability (Rycroft, 1972; Sonwalkar and Inan, 1989; Trakhtengerts, 1995; Singh et al., 1999b; Singh and Patel, 2004; Singh et al., 2007b).

**4.4.3.1 Hiss Emissions**

The spectral form of hiss emissions is a band limited non-dispersive signal. Its global distribution is characterized by three principal zones of intense activity, namely auroral hiss located around $70^0$ latitude, mid-latitude hiss near $50^0$ latitude and equatorial hiss observed below $30^0$ latitude. The intensity of the equatorial/low latitude hiss is less than those observed at middle and high latitudes (Hayakawa and Sazhin, 1992; Sazhin et al., 1993; Singh 1999; Singh et al., 1999b, 2001, 2007b). Recent observations clearly suggest that the equatorial region from low altitudes to the inner plasmasphere is an intense source of hiss emissions (Singh et al., 1999b; 2000, 2001; 2002; 2007b; Singh and Singh, 2002). Based on dynamic spectrum hiss emissions are classified in to continuous hiss and impulsive hiss. The occurrence probability of auroral hiss and equatorial hiss is increased when the $K_P$ index increases from 0 to 5 (Hayakawa et al., 1975; Singh and Singh, 2002). Impulsive hiss was reported during the expansion phase of a substorm in the midnight sector, whereas continuous hiss did not show any correlation with the local magnetic disturbances (Makita and Fukunishi, 1973). This shows that magnetospheric conditions control the generation and propagation of hiss emissions.

The early observations of hiss emissions were interpreted in terms of a wave-particle interaction mechanism where background noise could be amplified to the level of the observed hiss intensity (Solomon et al., 1988; Singh et al., 2001, 2007b). Santolik and Gurnett (2002) using spacecraft data concluded that hiss could arise from an extended sheet source. Santolik et al. (2001) analyzed high-rate waveform data of plasmaspheric hiss collected by the POLAR plasma wave instrument at high altitudes in the equatorial plasmasphere and suggested that waves with wave normals both parallel and anti-parallel to the geomagnetic field were generated through the gyroresonant interaction process with energetic electrons. Their amplification leads to the loss of resonantly interacting electrons from the radiation belts, particularly in the slot region $2 < L < 3$. However, typical wave growth rates inside the plasmasphere are not sufficient (Church and Thorne, 1983) to generate the observed intensity of hiss emissions (Singh et al., 2001).

Whistler mode waves propagating back and forth along geomagnetic field lines may undergo interference and diffraction processes to produce a hiss like spectrum (Dragonov et al., 1992; Bortnik et al., 2003 a,b). Sonwalkar and Inan (1989), analyzing

DE-1 satellite data, argued that the wave energy introduced in the magnetosphere by atmospheric lightning might play an important role in the embryonic generation of hiss emissions. Green et al. (2005) also discussed that hiss could be produced from lightning generated whistlers. However, the intensity distribution of hiss over the land mass and the ocean does not correlate with the distribution of lightning activity which shows stronger activity over the land mass than over the oceans (Green et al., 2005, 2006; Thorne et al., 2006). An extensive study showed that hiss intensity below 2 kHz has no correlation with land mass (Meredith et al., 2006).

Meredith et al. (2006) analyzed CRRES wave data together with the global distribution of lightning to test both the theories. They suggested that *in situ* amplification of wave turbulence in space seems to be the main source of wave power below 2 kHz, whereas wave power above 2 kHz is more likely to be related to lightning-generated whistlers. They further suggested that natural plasma turbulence should dominate the loss of relativistic (~ MeV) electrons in the slot region $2 < L < 3$. Meredith et al. (2007) estimated radiation belt electron loss time scales due to plasmaspheric hiss and lightning generated whistlers in the slot region. They showed that plasmaspheric hiss propagating at small and intermediate wave normal angles is a significant scattering agent in the slot region and beyond. In contrast, plasmaspheric hiss propagating at large wave normal angles and lightning generated whistlers do not contribute significantly to radiation belt loss. They further showed that plasmaspheric hiss may be an important loss process in the inner region of the outer radiation belt during magnetically disturbed periods.

Chorus emissions propagating long distances can also transform into hiss emissions. Santolik et al. (2006) using different plasma density models and reverse ray tracing computations showed that the observed low altitude hiss on the dayside at subauroral latitudes may have a possible source region near the geomagnetic equator at a radial distance between 5 and 7 Earth radii which is consistent with the source region of chorus. Thus they suggested that ELF hiss is nothing but chorus propagating to low altitudes, with their dynamic spectra being modified during propagation. Recently Bortnik et al. (2008) proposed a new explanation for the generation of plasmaspheric hiss that reproduces its fundamental properties. In this mechanism they argued that high intensity narrow band chorus can evolve into low intensity broadband noise filling the plasmasphere, ultimately being observed as hiss. Rodger and Clilverd (2008) commented that this modeling of hiss generation from chorus requires experimental confirmation. isThe connection between chorus and hiss is very interesting because chorus helps in the formation of high energy electrons (by acceleration) outside the plasmasphere (Horne et al., 2005) whereas hiss depletes these electrons at lower equatorial altitudes (Abel and Thorne, 1998).

**4.4.3.2 Chorus Emissions**

Chorus emissions are natural wave emissions generated by plasma instabilities in the Earth's magnetosphere observed first on the ground at middle and high latitudes (Helliwell, 1965). They occur in the frequency range from hundreds of Hz to several kHz and can have complex frequency spectra resembling a riser, faller, hook, inverted hook or even more complex structures on time scales of a fraction of second. Chorus is also observed at low latitude ground stations (Singh et al., 2000). Pickett et al. (2004) have reported remarkable cases of risers, fallers and hooks in the frequency range 1.5 - 3.5 kHz at and near the plasmapause observed on the four Cluster spacecraft wideband plasma wave receivers. Singh et al. (2004b) have also reported a rare observation of hisslers (LaBelle and Treumann, 2002) from a low latitude station, a quasi periodic falling noise which is feature of auroral broadband VLF hiss. Figure 5 shows the dynamic spectrum of chorus between 0.75 and 3.2 kHz recorded at Allahabad on 12 August 2007. Mid latitude chorus observed on the ground is correlated with X-ray bursts caused by 30 keV electrons (Rosenberg et al., 1990) and is mostly observed during disturbed period accompanied by hiss type emissions (Dowden, 1971).

Satellite observations have shown that chorus power fluxes increase outside the plasmasphere during enhanced geomagnetic activity (Meredith et al., 2001). However, Smith et al. (2004) showed that chorus power is initially depressed from the storm onset, reaching a minimum at about the time of the Dst minimum. This decrease in power on the ground surface may be due to propagation. In the recovery phase of the storm, chorus wave power rises above the prestorm level by as much as 40 dB.

The importance of chorus arises because these waves are believed to accelerate electrons outside the plasmasphere. Studies have shown that relativistic electron enhancements are associated with elevated fluxes of lower energy electrons and prolonged periods of enhanced chorus amplitudes lasting for few days (Meredith et al., 2002, 2003; Miyoshi et al., 2003). The observations of a local peak in phase space density (Miyoshi et al., 2003), flat top pitch angle distributions (Horne et al., 2003) and energy dependence in particle spectrum (Summers et al., 2002) support the idea that electrons are accelerated by the chorus emissions during wave (chorus) – particle (electron) interactions.

The other effect of chorus is to cause burst precipitation (Horne and Thorne, 2003) and the depletion of radiation belt electrons. These precipitated energetic electrons create additional ionization in the lower atmosphere (Rodger et al., 2007) and cause the absorption of whistler mode waves as they propagate through the medium (Smith et al., 2004). Rodger et al. (2007) computed the energy spectra of precipitated electrons based on current models of chorus propagation and wave-particle interaction theory. They showed that the results are not consistent with the experimentally observed radio wave perturbations. Chorus can act as a mediating agent transferring energy from the lower energy (~ 10 – 100 keV) electrons which are predominantly precipitated to the relativistic (~ MeV) electrons which are accelerated (Meridith et al., 2002; Horne et al., 2005; Bortnik et al., 2007).

The generation mechanism of chorus has been extensively studied in the past (Sazhin and Hayakawa, 1992 and references theirin). The nonlinear cyclotron resonance interaction between whistler mode waves and counter streaming electrons is most widely used theory (Trakhtengerts, 1999; Singh et al., 2000; Singh and Patel, 2004; Titova et al., 2003). During the development of the cyclotron instability, a singularity in the form of the step on the distribution function of the energetic electrons is formed at the boundary in velocity space between resonant and non-resonant electrons, and the backward wave oscillator (BWO) regime is realized; discrete emissions such as chorus are generated. Singh and Patel (2004) used the BWO mechanism to explain some features of chorus observed at the Indian Antarctic Station, Maitri.

Chorus is believed to be generated near the geomagnetic equator, where the first derivative of the magnetic field strength along the field line is nearly zero (Helliwell, 1967; Lauben et al., 2002). The analysis of multi satellite Cluster data of chorus emissions (Parrot et al., 2003; Santolik et al., 2003; 2005; 2006) show that the generation region is localized near the equatorial cross-section of the L~4 magnetic flux tube, and has a typical scale size of ~ 2000 km along the magnetic field lines. The source center is defined by the balance of the Poynting flux parallel and antiparallel to the field line (Santolik et al., 2005; Santolik, 2008). The central position is found to move randomly back and forth over a few thousand of kilometers on time scales of minutes. The chorus waves propagating along geomagnetic field lines may be reflected back to the source region (Parrot et al., 2004; 2006). The intensity ratio between the magnetic components of waves coming directly from the equator and waves returning to the equator was observed to be between 0.005 and 0.01 (Santolik, 2008).

Ray tracing results show that the chorus waves are generated with oblique wave vectors pointing towards the Earth in an equatorial region between 5 and 7 Earth radii from the Earth (Chum and Santolik, 2005). Moreover, the wave normals are nearly field aligned when the waves again cross the equator inside the plasmasphere which is consistent with the observed wave normal of plasmaspheric hiss and which makes further amplification of these waves possible (Santolik et al., 2001). This also shows that waves generated with finite wave normal angle can, after one/two reflections, have wave normals almost parallel to the geomagnetic field line and so could be received on the Earth's surface (Singh et al., 2000; Singh and Patel, 2004).

Trakhtengerts et al. (2007) analyzed Cluster data obtained on two different geomagnetically active days of April 18, 2002 and March 31, 2001 and showed that the frequency spectrum of individual chorus elements depends on the position of the observation point in and near the generation region. Breneman et al. (2007) reported correlated chorus elements with different frequency/time characteristics as seen on the four different Cluster spacecraft. They used a cross-correlation analysis to quantify the dispersive time delay between each frequency of a chorus element as it arrives at Cluster spacecraft pairs; this is compared with ray-tracing results in order to identify source locations that are consistent with the observed delays. Chum et al. (2007) have shown that a quasi-stationary, or static, source that varies in time can reproduce chorus observations.

Satellite data, showing internal fine structure of each wave packet consisting of a sequence of separate sub-packets, has been explained by the generation of sidebands during the evolution of a chorus wave packet and by the beating effect of simultaneously present signals at closely separated frequencies (Nunn et al., 2005). Waveform analysis shows that the duration of sub-packets is variable from a few milliseconds to a few tens of milliseconds (Santolik et al., 2003; 2006; Santolik, 2008).

Systematic observations have confirmed that in most cases the amplitudes of chorus emissions are indeed sufficiently high to be governed by nonlinear effects. The measured amplitudes of the sub-packets reached more than 30 mV/m (Santolik, 2008). Cattell et al. (2008) reported maximum amplitude of dawnside chorus to be 240 mV/m from the S/WAVES instrument onboard STEREO spacecraft. The high amplitudes and fine structures clearly suggest that nonlinear effects play an important role in the microphysics governing the interaction of these waves with charged particles populating the interaction region in particular and the magnetosphere in general.

### 4.4.3.3 Periodic and Quasiperiodic Emissions

VLF emissions of short bursts repeated at regular intervals of the order of few seconds termed periodic emissions are observed both on the ground as well as onboard satellites. They are classified as either dispersive or as non-dispersive. In the dispersive type the period between bursts varies systematically with frequency whereas in the non dispersive type there is a little or no observable systematic change in period with frequency (Helliwell, 1965; Sazhin and Hayakawa, 1994). Considering the periodicity to be associated with the generation process, two models were suggested, one involving the bounce period of the charged particles and the other involving the two-hop whistler transit time (Dowden and Helliwell, 1962).

Quasi periodic emissions (QP) are VLF emissions with periods of 20-50 s (Helliwell, 1965). Kitamura et al. (1968) were the first to subdivide QP emissions into those which were associated with corresponding geomagnetic pulsation activity (QP1), and those which were not associated with pulsation activity (QP2). Both QP1 and QP2 were further classified according to the form of their spectrograms (Sazhin and Hayakawa, 1994 and references therein). Smith et al. (1998) analyzed data from South

Pole, Halley and found a strong association between periodic emissions (PE) and the simultaneous appearance of QP2. Incorporating a larger data set from a latitudinal array of Antarctic stations, Engebretson et al. (2004) studied latitudinal and seasonal variations of quasi-periodic and periodic emissions. PEs and QPs of type I was found to have different latitudinal, seasonal and diurnal occurrence patterns. PEs occurred more around $60^0$ geomagnetic latitude whereas QPEs occurred around $65^0$-$70^0$ geomagnetic latitude. PEs occurred at all local times but the diurnal variation was latitude dependent. PEs were more common during the months of May to September whereas QPEs were observed during the months of October to March. Pasmanik et al. (2004) presented the results of a case study of quasi-periodic (QP) ELF/VLF hiss emissions detected on board the Freja and Magion 5 satellites. They reported an event with an increase in the frequency drift rate during the generation of a single element of QP elements having different frequency drift rates. They have also shown that the generation of QP emissions can be accompanied by the generation of discrete emissions.

### 4.4.3.4 Triggered Emissions

Triggered emissions exhibit a bewildering variety of dynamic spectral forms (Helliwell, 1965; Nagano et al., 1996; Nunn et al., 1997; Smith and Nunn, 1998; Singh et al., 2003a; Singh and Patel, 2004; Siingh et al., 2005b) and follow a source which could be a whistler (Storey, 1953; Nunn and Smith, 1996), discrete emissions (Helliwell, 1965), signals from VLF transmitters (Helliwell, 1965; Bell et al., 1982), power line radiation from the world's power grids (Helliwell et al., 1975; Park and Change, 1978; Luette et al., 1979) and the upper frequency boundary of hiss (Helliwell, 1969; Reeve and Rycroft, 1976 a,b; Koons, 1981; Hattori et al., 1989, 1991; Singh et al., 2000; Singh and Ronnmark, 2004). The observations in general support the idea that strong VLF emissions may be triggered by a very weak signal. Perhaps even sometimes the triggering source may be invisible. The dynamic spectra of triggered emissions are complex and the complexity of the problem can be appreciated from the fact that (a) triggered emissions are non-stationary/transient phenomena, (b) the variation of frequency varies widely from event to event and (c) the frequency of these emissions can differ considerably from that of the triggering signals.

A number of nonlinear processes have been suggested to explain the phenomena (Helliwell, 1967, Omura et al., 1991; Nunn and Smith, 1996; Smith and Nunn, 1998; Hobara et al., 1998). The nonlinear effect can be categorized into wave-wave interactions (Harker and Crawford, 1969) and wave-particle interactions. There seems to be general agreement that the mechanism responsible for the generation of triggered emissions is a nonlinear interaction between a finite amplitude wave-train and the particles that happen to be in resonance with it. The nonlinear interaction produces phase bunching of the resonant and nearly resonant electrons, thus giving rise to a current, which acts like an antenna and generate VLF waves/emissions (Helliwell, 1967; Dysthe, 1971).

The second order resonance condition close to the geomagnetic equator in the magnetosphere is valid for this slowly varying inhomogeneous interaction region, and determines the frequency spectrum of the discrete emissions generated by the energetic electron beam. Roux and Pellat (1978) discussed a self sustaining theory in which particles which are detrapped at the triggering wave termination are phase organized and act coherently for a while they therefore give rise to an emission with either falling or rising frequency, depending upon the sign of the inhomogeneity variation. Further, the spatially averaged part of the distribution function of these suddenly detrapped electrons generates an instability, which amplifies the emitted waves. Helliwell and Inan (1982) have discussed a feedback model in which the interaction region centered on the magnetic equator is treated like an unstable feedback amplifier. Molvig et al. (1988) have developed a self-consistent theory of triggered whistler emissions, which is capable of predicting the observed dynamic spectrum of the emissions. In the feedback model constant frequency oscillations are generated on the equator, risers and fallers are generated on the downstream and upstream sides of the magnetic equator, respectively (Helliwell, 1967; Helliwell and Inan, 1982).

Pickett et al. (2004) reported multipoint Cluster observations of remarkable cases of triggered emissions of VLF risers, fallers and hooks in the frequency range of 1.5 to 3.5 kHz with a frequency drift for the risers on the order of 1 kHz/s. These emissions appear to be triggered out of the background hiss. Trakhtengerts et al. (2003) considered the effect of initial phase bunching of energetic electrons on the generation of triggered ELF/VLF emissions in the magnetosphere. They showed that the electrons interacted with the primary whistler wave packet to form a phase-bunched beam in velocity space, which serves as a traveling wave antenna.

**4.4.3.5 VLF Waves as Magnetospheric Probes**

Waves propagating through different regions of space carry information about the medium through which they travel. By analyzing the received wave features, it is possible to derive information about the medium such as the electron and proton density, temperature, and electric and magnetic field distributions in the medium (Helliwell, 1965; Singh et al., 1998 a,b, and references therein ). Apart from these diagnostic features, these waves carry thunderstorm energy from the lower atmosphere to the plasmasphere and magnetosphere and thus couple the atmosphere to the plasmasphere and magnetosphere (Siingh et al., 2005a).

The dispersion property of whistlers is widely used to yield information about the medium parameters such as electron density and total electron content of a flux tube (Sazhin et al., 1992; Singh, 1993; Singh et al., 1993; Singh and Singh, 1997; Singh et al., 1998a), and electron temperature (Scarf, 1962; Guthart, 1965; Sazhin et al., 1990, 1993). Similarly the analysis of ion whistlers provides information about ion density and ion temperature (Singh et al., 1998b). During continuous whistler monitoring if one finds changes of nose frequency with time, it is possible to estimate the large scale electric fields (Bernard, 1973; Block and Carpenter, 1974; Park, 1976, 1978; Singh, 1993) present in the magnetosphere. Coupling of the ionosphere and protonosphere was also studied using whistler data (Park, 1978; Lalmani et al., 1992; Singh, 1993). One of the important findings of the whistler studies was the discovery of the plasmapause where the electron density suddenly decreases by an order of magnitude or more within a fraction of an Earth's radius (Park and Carpenter, 1978). In low latitude regions non-nose whistlers are observed; using the extension method the full dynamic spectra can be constructed to determine the path of propagation (Singh et al., 2006b).

The electron density, electron temperature and other parameters derived from whistler measurements compare well with direct rocket/satellite measurements (Park et al., 1978; Singh et al., 1998 a, b). The study of the latitudinal and longitudinal distribution of electron density and its long term variations using rockets/satellites is financially and technically challenging, whereas these can be studied very readily by the whistler technique at a number of stations spread in latitude and longitude. Singh et al. (2006a) analyzed diffuse whistlers recorded at the low latitude station Varanasi, India, on January 11, 1998 using matched filtering technique to estimate parameters of the medium. They showed that the diffuse whistlers are formed due to merging of a large number of fine structure traces.

The remote sensing method based on ground-based measurements provides average information and is model dependent. It does not yield the local value of the electron density in the magnetosphere. Trakhtengerts and Rycroft (1998) suggested a method based on the phenomena of nonlinear whistler wave reflection from the lower hybrid resonance level (Trakhtengerts et al., 1996). This allows one not only to measure the local value of electron density but also localizes the place of measurement in the magnetosphere. Using this method, one can obtain important information about short scale low frequency turbulence (ion-cyclotron waves) present in the magnetosphere. Considering the upper cut-off frequency of the nose whistler to be due to thermal attenuation, the electron temperature has been estimated to vary between 1.7 eV ($2 \times 10^6$ K) and 4 eV at L = 4. The method was not widely used because it is difficult to distinguish whether the upper cut-off in the spectrum was due to thermal attenuation or to propagation effects (Sazhin et al., 1990). Attempts have also been made to infer magnetospheric electron temperature from whistler dispersion measurements (Sazhin et al., 1993).

The diffuse nature of whistler traces yields information about duct width (~100 km). Duct life time usually varies between 30 minutes and 2 hours, although duct life times as high as 1-2 days have been reported (Singh et al., 1998a; Singh and Singh, 1999). Ducts occupy a relatively small volume (0.01%) in the magnetosphere. The electric field from a lightning discharge penetrates the ionosphere and can create a density channel in the magnetosphere along geomagnetic field lines. Park and Dejnakarintra (1973) computed high altitude electric fields due to thunderclouds taking into account the presence of ionosphere. They showed that the flux interchange mechanism of duct formation is plausible. Rodger et al. (1998) have shown that under 'typical' atmospheric conductivity conditions the high altitude electric fields from even giant thunderclouds are too small to create a realistic whistler duct in a realistic period, during both day and night conditions.

**4.5 HF, VHF and UHF Waves**

VHF and UHF radiation from lightning yields rich information about the morphology of separate micro-discharges, pulse trains and bursts, their connection with different phases of a lightning flash, and individual and averaged frequency spectra of radiation from the discharge (Rakov and Uman, 2003). In fact, the elementary micro discharges with the front duration ~ 10 ns and complete duration ~ 1 μs (Rakov and Uman, 2003) are the elementary emitters in the VHF/UHF band. The contribution of the return stroke in this very high frequency range is small (Rakov and Uman, 2003).

Hayakawa et al. (2008) have developed a three dimensional simulation of micro-discharge activity in thunderstorm clouds. They showed that the simulated waveforms are close to those observed and the temporal development, with the duration of pulse trains from tens to hundreds of microseconds and the micro-discharge number rate being in agreement with the corresponding experimental data. During the Euro-sprite campaign high frequency (3 – 30 MHz) electromagnetic signals were recorded at four stations in France (Neubert et al., 2008) and were correlated with cloud-to-ground discharges and sprites; they reported weak high frequency emissions during the time interval when sprites were observed.

VHF signals were also observed onboard the ALEXIS satellite in the form of pulse pairs, called as Trans-Ionospheric Pulse Pairs (TIPPs) (Holden et al., 1995; Mossy and Holden, 1995). The dynamic spectra of these signals resemble whistlers, each with a uration of a few microseconds separated by some tens of microseconds. Holden et al. (1995) proposed that TIPPs are generated during electrical activity of clouds and the dispersion is accounted for by the signal passing through the ionosphere. It was proposed that the first signal could be associated to the cloud-to-ground discharge and the second with the cloud-to-ionosphere discharge (Roussel-Dupre and Gurevich, 1996). The observed bifurcation at lower frequencies was attributed to the propagation of both ordinary and extraordinary modes (Holden et al., 1995; Singh and Singh, 2005b). Representing the CG and high altitude discharge currents by a combination of Dirac delta functions separated in the time domain, Singh and Singh (2005b) simulated the dynamic spectra of the TIPPs. Details of simulation technique are given by Singh and Singh (2005a). The spectrograms of TIPPs depend on the high altitude current, the time lag of the high altitude discharge and the electron density of the $F_2$ layer.

## 5. Summary and Conclusions

In this paper, we have summarized certain aspects of electrical discharges in thunderstorms and associated phenomena. Both the evolution and electrification of thunderstorms are briefly discussed. However, the microphysics of charge separation in thunderclouds under different meteorological conditions is not fully understood. This requires further investigations using field and laboratory data and three dimensional simulations of the microphysical processes involved.

The bidirectional (cloud-to-ground and cloud-to-high altitude) discharges from thunderclouds lead to many spectacular phenomena, and represent an energy source in a planetary atmosphere coupling different layers to the outermost regions (ionosphere, magnetosphere). This is still an active research field, especially on discharges in the mesosphere. Various optical bands in the mesospheric discharges (TLEs) have been observed, but distribution of intensities as a function of wavelength is not precisely known. Further experimental measurements and modeling work is required. The observations of sprites have revealed that many processes are involved. Observations show an asymmetry in cloud-to-ground discharges which generate sprites; negative CG discharges are rarely associated with sprites. Further model computations could show that sprite sizes are connected with the lightning's charge moment change.

The infrasound observations show that the energy input during sprites is quite high, ~ 0.4 – 40 GJ (Farges et al., 2005), and this energy would produce perturbations in the atmosphere. This speculation has to be explored further. The infrasound can propagate to long distances and hence the perturbations might also be over a wide area.

The upward discharge creates a conducting path in the upper atmosphere so that the role of sprites and TLEs in the global electric circuit becomes important. The electric fields established in the mesosphere following a lightning discharge are not properly understood, and the relaxation time scale of quasi-electrostatic fields and its dependence on the ambient parameters of the medium are not well known. A complete understanding of the electric fields generated during intra-cloud, cloud-to-cloud, cloud-to-ground and cloud-to-ionosphere discharges remain a problem which requires both experimental and theoretical modeling efforts.

The electrical processes acting in thunderstorms control the flow of vertical current, which near the cloud surface create space charges which rapidly attach to droplets, aerosol particles and ice forming nuclei and affect storm dynamics. The charging time constant ranges from minutes to hours, which are comparable to typical convection and turbulence characteristics times. Therefore, time dependent cloud charging models, including turbulence and convection, are required for detailed study.

Upward discharges and their associated electromagnetic radiation modify mesospheric and stratospheric processes by changing the concentration of different constituents such as $NO_x$ and $HO_x$. Streamers in sprites produce heat in the neutral atmosphere and aid in the generation of $NO_x$. Thus perturbations in mesospheric chemical constituents require further study. TLE discharges in the mesosphere provide an opportunity to probe this region which remains inaccessible except using rockets. Hence in-depth studies of TLEs become more relevant and important.

Processes involving nonlinear behavior, e.g., wave-particle interactions and wave-wave interactions, still require further studies. The role of space weather parameters on the generation and propagation of ELF/VLF emissions have not been comprehensively studied. Their fine structure and other morphological features observed by various satellites, including Cluster, is an important area to be explored. ELF waves have been widely used to monitor the global occurrence of lightning and sprites. However, the mechanism by which sprites generate ELF waves is not properly known and hence further study in this area is suggested.

The thunderstorm electrification process and downward lightning discharges play a significant role in the Earth's climate. They also couple the troposphere to higher regions of the atmosphere and to the ionosphere and magnetosphere. Many complex processes are involved which warrant detailed study.

**Acknowledgements**

DS acknowledges financial support from the Ministry of Earth Sciences (MoES), Government of India, New Delhi and also Head, I&OT Division for support. This work is partly supported by DST, New Delhi under SERC project and partly by ISRO, Bangalore, under the CAWSES program. R. P. Patel thanks DST for the award of a FASTTRACK fellowship (SR/FTP/PS-12/2006). The authors thank both the anonymous reviewers for their critical comments which helped in improving the scientific value of this paper. They also express their gratitude to Prof. M. J. Rycroft for his valuable suggestions.

# Captions to Figures:

Fig. 1. Various currents that flow in the vicinity of an active thundercloud; there are five contributions of the total current. In the same figure are shown the various Transient Luminious Emissions (TLEs) of the stratosphere and mesosphere

Fig. 2. Schematic of a section through the global atmospheric electric circuit in the dawn–dusk magnetic meridian (Tinsley, 2008). The tropospheric and stratospheric column resistances at a given location are represented by $T$ and $S$, with subscripts referring to equatorial, low, middle, high, and polar latitudes. The geometry is essentially plane-parallel, with only small changes in $T$ due to changes in cosmic ray fluxes at low latitudes, but large changes in $T$ and $S$ due to cosmic ray and other energetic space particle fluxes at high latitudes. The variable solar wind generators affect $V_i$ at high latitudes, and the variable highly electrified cloud generators, represented by the generator symbol in the equatorial plane, affect $V_i$ globally, and volcanic aerosols as well as the energetic particles affect $T$ and $S$; with all acting together to modulate the ionosphere-Earth current density $J_z$.

Fig. 3. (a) Waveform, (b) spectrogram, and (c) power spectrum of a sferic recorded on 23 March, 2008 at Allahabad (India).

Fig. 4. Dynamic spectra of four whistlers recorded at Allahabad, (India) a low latitude Indian station on 17 June, 2008 at 12:40 UT.

Fig. 5. Two typical spectrogram of chorus emissions recorded at an Indian ground station Allahabad (India) on 12 August, 2007 at 19:30 UT.

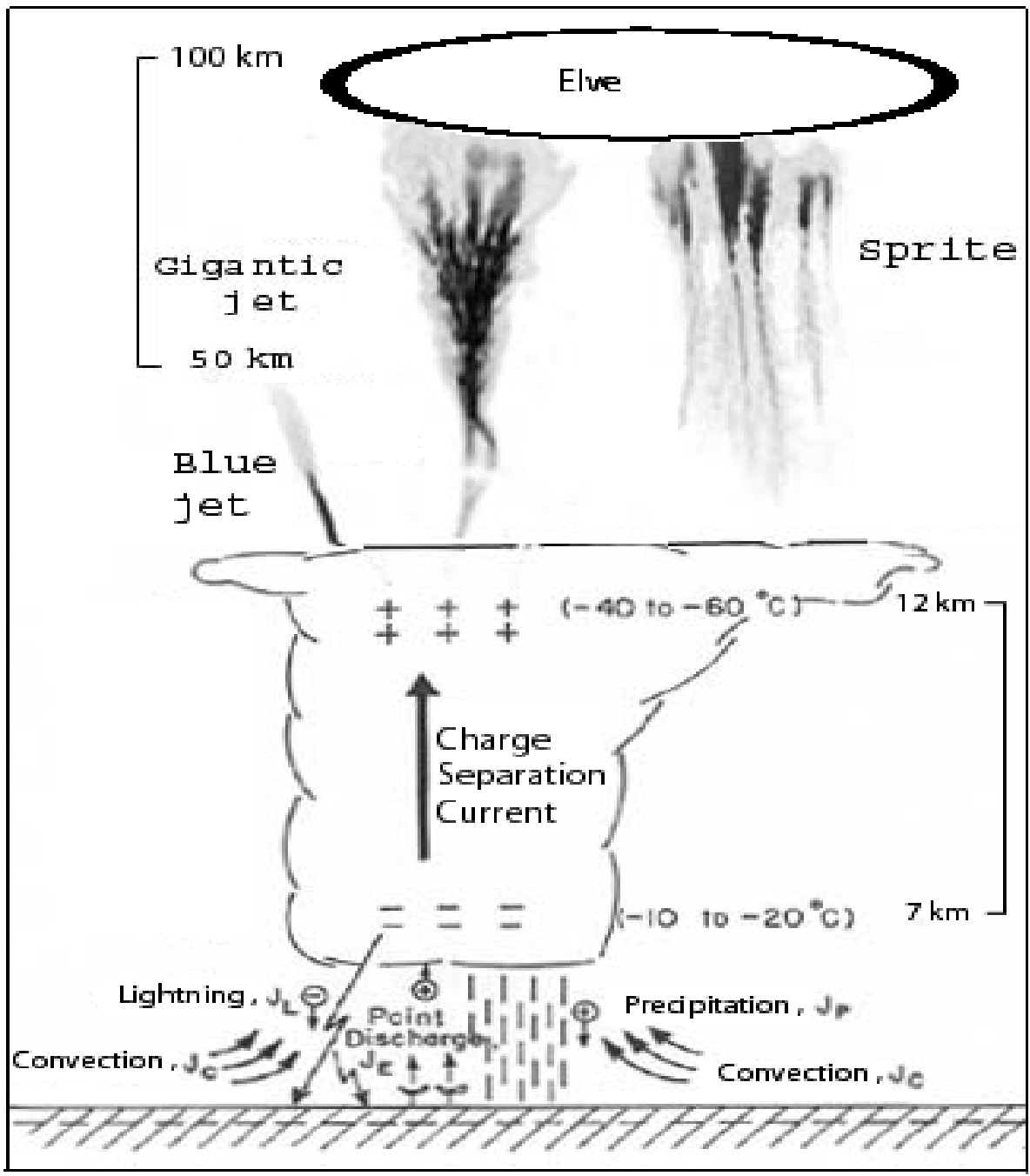

Figure-1

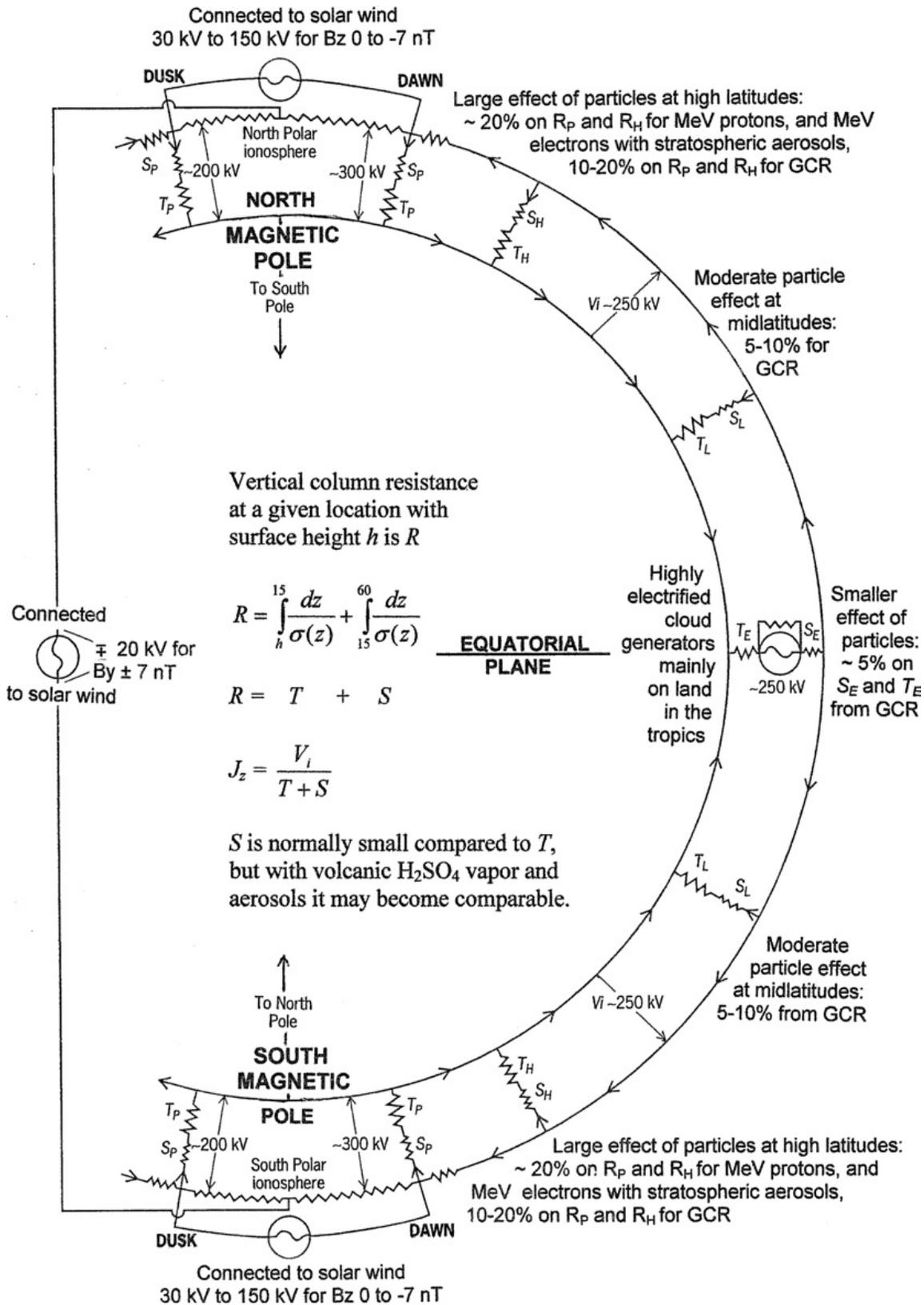

Figure-2

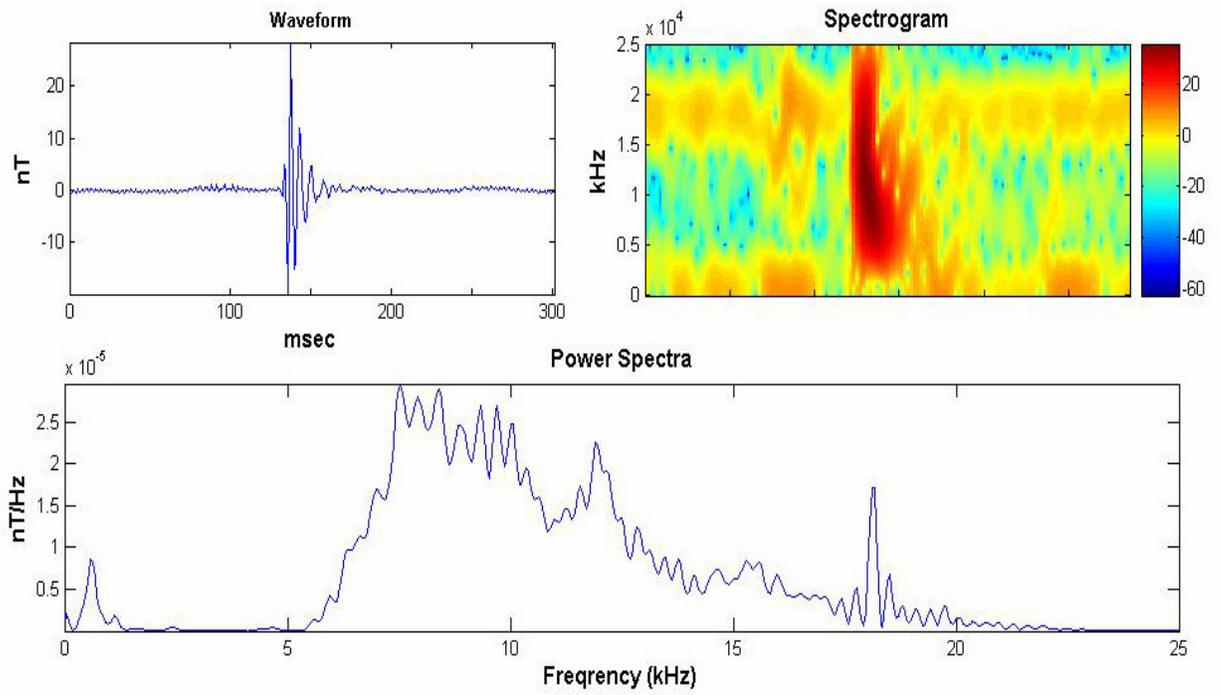

Figure-3

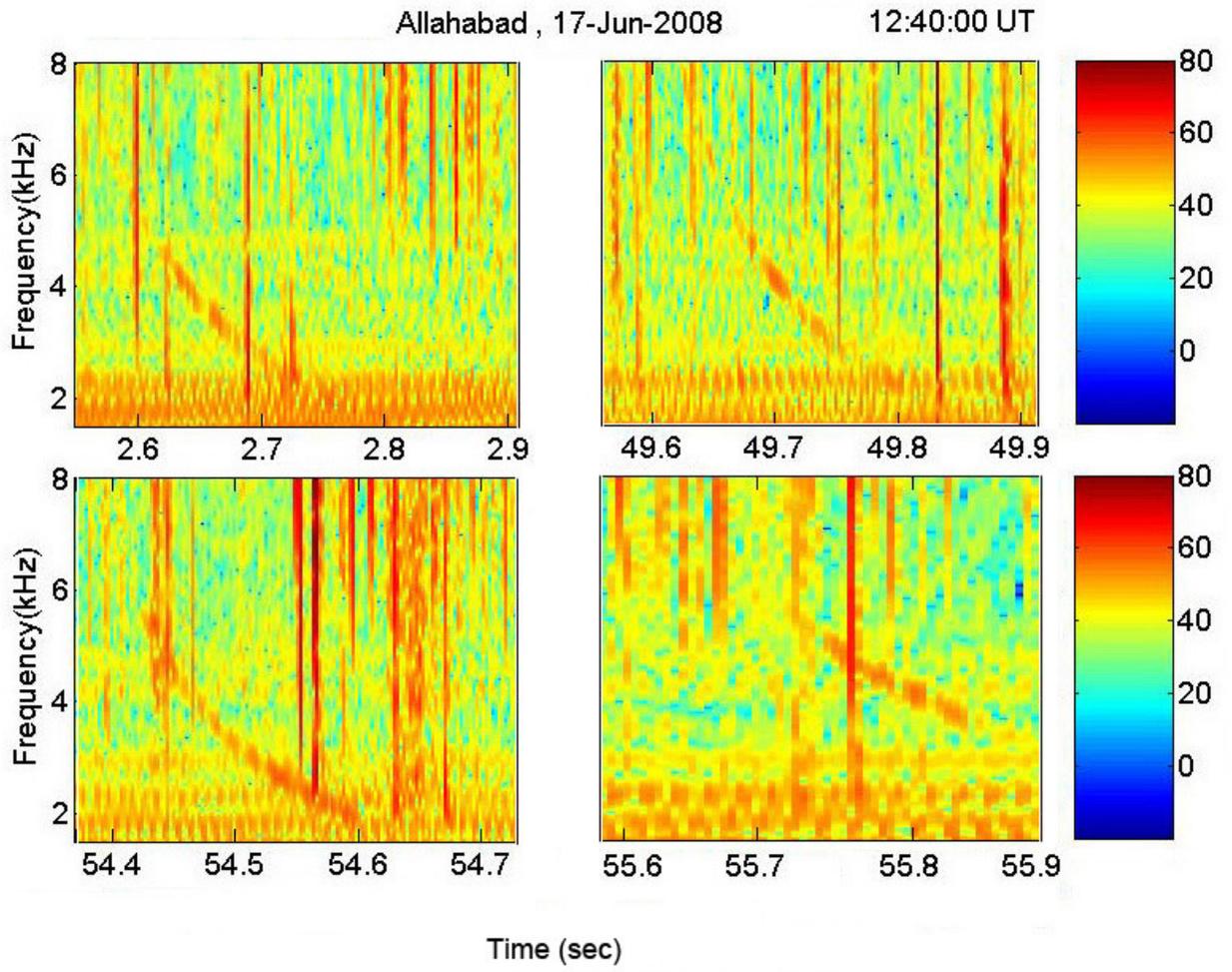

Figure-4

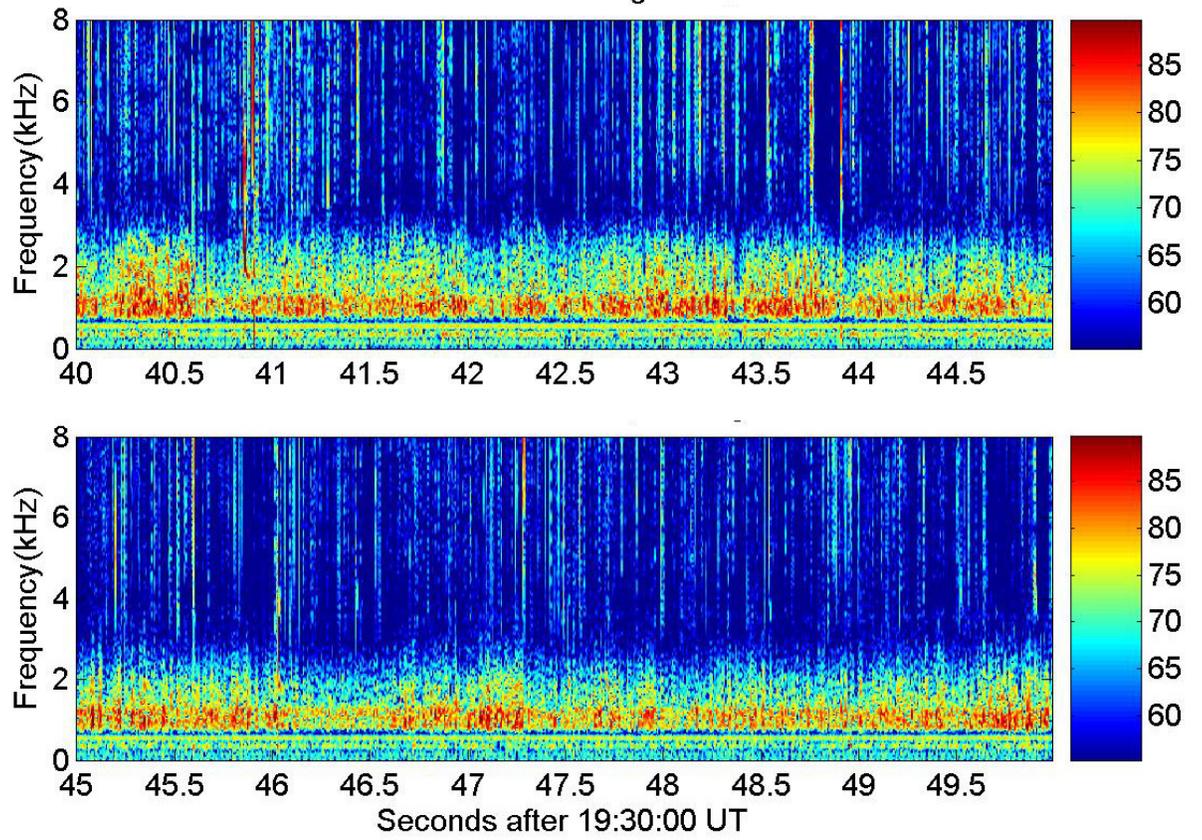

Figure-5